\documentclass[reprint,showpacs,showkeys,amsmath,amssymb,superscriptaddress,aps,pra]{revtex4-1}

\usepackage{graphicx}
\usepackage{float}
\usepackage{lipsum}
\usepackage{epstopdf}
\usepackage{bm}
\usepackage{amssymb}
\usepackage{amsmath}
\usepackage{bbold}
\usepackage[colorlinks=true, linkcolor=blue]{hyperref}
\usepackage{ulem}
\usepackage{color, soul}
\usepackage{xcolor}
\usepackage[utf8]{inputenc}
\DeclareUnicodeCharacter{2212}{-}
\DeclareUnicodeCharacter{B0}{\textdegree}
\usepackage{chemformula}
\DeclareUnicodeCharacter{0301}{\'{e}}
\usepackage{gensymb}
\usepackage{graphicx}
\usepackage{textcomp, gensymb}
\usepackage{tikz}
\usepackage{makecell}
\usepackage{float}
\usepackage{silence}
\usepackage{multirow}

\DeclareUnicodeCharacter{0301}{\'{e}}
\DeclareUnicodeCharacter{2218}{\'{e}}
\DeclareUnicodeCharacter{2082}{\'{e}}

\begin{document}

\title{Unconventional magnetoelastic behavior in Al-rich Co–Fe–Al Films with inverse-Heusler-like local order for flexible spintronics}

\author{Rupalipriyadarsini Chhatoi$^\ddagger$}
\affiliation{Laboratory for Nanomagnetism and Magnetic Materials (LNMM), School of Physical Sciences, National Institute of Science Education and Research (NISER), Jatni, Odisha 752050, India}

\affiliation{Homi Bhabha National Institute, Training School Complex, Anushaktinagar, Mumbai 400094, India}
\altaffiliation{These authors contributed equally to this work}
\author{Anuroopa Behatha$^\ddagger
$}
\affiliation{Indo-Korea Science and Technology Center (IKST), Bangalore, India}
\altaffiliation{These authors contributed equally to this work}
\author{Swayang Priya Mahanta}
\affiliation{Laboratory for Nanomagnetism and Magnetic Materials (LNMM), School of Physical Sciences, National Institute of Science Education and Research (NISER), Jatni, Odisha 752050, India}

\affiliation{Homi Bhabha National Institute, Training School Complex, Anushaktinagar, Mumbai 400094, India}
\author{Shubhransu Sahoo}
\affiliation{Laboratory for Nanomagnetism and Magnetic Materials (LNMM), School of Physical Sciences, National Institute of Science Education and Research (NISER), Jatni, Odisha 752050, India}

\affiliation{Homi Bhabha National Institute, Training School Complex, Anushaktinagar, Mumbai 400094, India}
\author{Soubhagya Dash}
\affiliation{Laboratory for Nanomagnetism and Magnetic Materials (LNMM), School of Physical Sciences, National Institute of Science Education and Research (NISER), Jatni, Odisha 752050, India}

\affiliation{Homi Bhabha National Institute, Training School Complex, Anushaktinagar, Mumbai 400094, India}
\author{Bhuvneshwari Sharma}
\affiliation{Laboratory for Nanomagnetism and Magnetic Materials (LNMM), School of Physical Sciences, National Institute of Science Education and Research (NISER), Jatni, Odisha 752050, India}

\affiliation{Homi Bhabha National Institute, Training School Complex, Anushaktinagar, Mumbai 400094, India}
\author{Abhisek Mishra}
\affiliation{Laboratory for Nanomagnetism and Magnetic Materials (LNMM), School of Physical Sciences, National Institute of Science Education and Research (NISER), Jatni, Odisha 752050, India}

\affiliation{Homi Bhabha National Institute, Training School Complex, Anushaktinagar, Mumbai 400094, India}
\author{Esita Pandey}
\affiliation{Laboratory for Nanomagnetism and Magnetic Materials (LNMM), School of Physical Sciences, National Institute of Science Education and Research (NISER), Jatni, Odisha 752050, India}

\affiliation{Homi Bhabha National Institute, Training School Complex, Anushaktinagar, Mumbai 400094, India}
\author{Satadeep Bhattacharjee}
\email{s.bhattacharjee@ikst.res.in}
\affiliation{Indo-Korea Science and Technology Center (IKST), Bangalore, India}
\author{Subhankar Bedanta}
\email{sbedanta@niser.ac.in}
\affiliation{Laboratory for Nanomagnetism and Magnetic Materials (LNMM), School of Physical Sciences, National Institute of Science Education and Research (NISER), Jatni, Odisha 752050, India}
\affiliation{Homi Bhabha National Institute, Training School Complex, Anushaktinagar, Mumbai 400094, India}
\affiliation{Center for Interdisciplinary Sciences (CIS), National Institute of Science Education and Research (NISER)}
\begin{abstract}

Strain engineering of magnetic properties offers a promising route toward flexible spintronic applications. Here, we report an unconventional magnetoelastic response in Al-rich Co–Fe–Al thin films on flexible substrates using strain-dependent magneto-optical Kerr effect microscopy and magnetometry measurements. While the films exhibit a conventional positive magnetostriction coefficient, consistent with standard in-plane easy-axis rotation under stress, their saturation magnetization increases under compressive strain and decreases under  tensile strain. First-principles calculations reveal that this unconventional response originates from a strain-induced competition between exchange splitting and crystal-field effects  in an inverse-Heusler-like local environment created by Al enrichment. This leads to a highly sensitive, sublattice-dependent magnetic state, consistent with a strain-induced reconfiguration of Co and Fe moments. Our results demonstrate that local compositional tuning can fundamentally alter magnetoelastic behavior, establishing strain-controlled sublattice compensation as a route toward programmable magnetic functionality in flexible spintronic systems.

\end{abstract}


\maketitle

Flexible spintronic platforms enable strain-mediated control of magnetic properties by integrating functional magnetic materials on deformable substrates\cite{corzo2020flexible,hwang2012flexible,han2013layer}. In such systems, epitaxial
mismatch, substrate bending, or piezoelectric actuation
can impose tensile or compressive strain, which modifies magnetic anisotropy, coercivity, and magnetization through magnetoelastic coupling\cite{schafer2009suppression,avc2008influence,sheng2018flexible,yang2024advances,chen2021mechanically,zhao2018low,emori2017coexistence,finizio2014magnetic,begue2021strain,heuver2015strain,Liu_2013}. Identifying material systems with robust and tunable magnetic responses under strain remains a key requirement. 

Heusler alloys, are attractive candidates due to their high Curie temperature, sizable spin polarization, and tunable magnetic anisotropy\cite{wen2019spin,wurmehl2006investigation,chumak2021magnetoelastic,zhang2018direct,titov2018microstructure}. The full Heusler compound Co$_2$FeAl, in its regular $L2_1$ structure, has been extensively investigated for magnetic tunnel junctions, spin-transfer torque, and spin-orbit torque applications \cite{kudryavtsev2007evolution,zhang2018direct,titov2018microstructure}. More recently, inverse-Heusler derivatives such as Fe$_2$CoAl with XA-type ordering have also
attracted attention, as they offer modified local coordination and electronic structure while preserving the overall Co–Fe–Al chemistry\cite{ahmad2020competition,rai2020pressure,siakeng2018electronic}. Strain is a natural control knob in thin-film Heusler systems. In the conventional picture, strain modifies magnetism through electronic band effects, where compressive strain broadens the d bands and reduces local magnetic moments, whereas tensile strain narrows the bands and enhances exchange splitting, leading to increased magnetization \cite{ahmad2020phase,martinez2026tailoring,Oh2026}
Most studies on flexible Heusler systems have focused on compositions close to stoichiometric Co$_2$FeAl, where this behavior is generally preserved.
\par
In this work, we deliberately investigate Al-rich Co–Fe–Al thin films, where controlled Al enrichment significantly modifies the local chemical environment of Fe and Co atoms and alters the balance between intra-atomic exchange and crystal-field splitting. In this regime, the strain response is no longer trivial, as changes in local coordination and bond lengths can stabilize competing spin configurations. As a result, the sign of the magnetoelastic response itself may be altered. We demonstrate that Al-rich films (Al $\sim$ 52 at.\%) grown on flexible polyimide substrates exhibit a reversible strain-controlled rotation of the magnetic easy axis, and within the elastic limit, the magnetostriction value remains positive. More importantly, the saturation magnetization shows an inverted strain dependence, i.e., it increases under compressive strain and decreases under tensile strain, which is opposite to the behavior commonly reported in Heusler alloys \cite{ahmad2020phase,martinez2026tailoring,Oh2026}. First-principles calculations reveal that this unconventional behavior originates from an inverse-Heusler–like local environment, where strain modifies $d-d$ hybridization and crystal-field splitting, driving a reconfiguration of Fe and Co sub-lattice moments. Our findings establish Al-rich Co–Fe–Al films as a platform for strain-programmable magnetism, demonstrating that compositional tuning can qualitatively reconfigure magnetoelastic response in Heusler systems.
\\
\begin{figure*}[!htbp]
\centering
\includegraphics[width=1\textwidth]{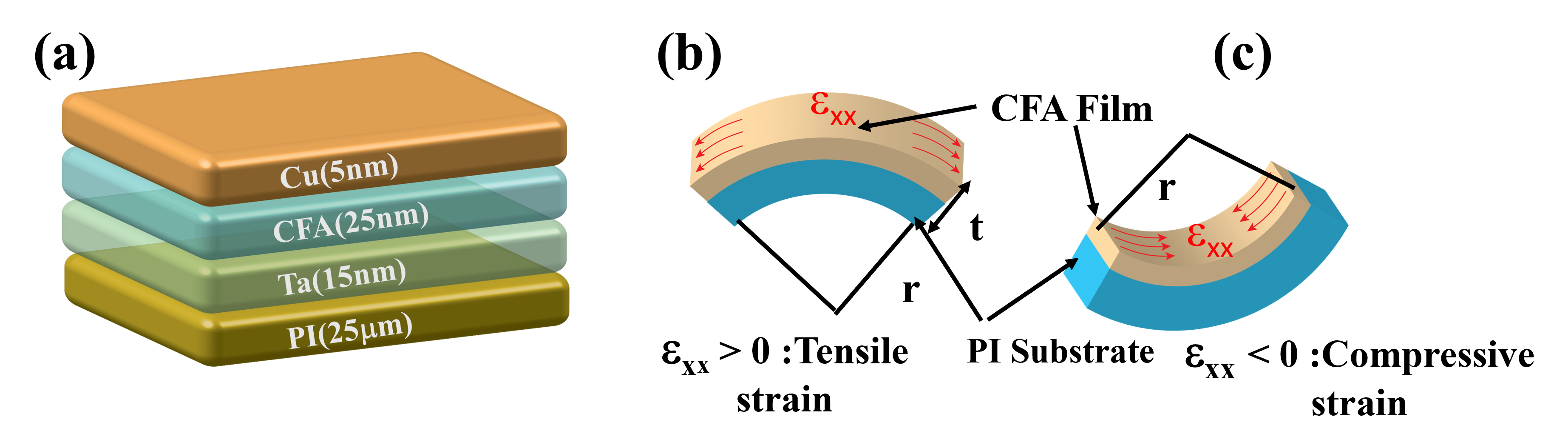}
\caption {Schematic illustration of the CFA thin-film stack and the generation of uniaxial in-plane strain by bending the flexible PI substrate. (a) The film stack consists of $\mathrm{PI}\;(25~\mu\mathrm{m})$/Ta (15 nm)/CFA (25 nm)/Cu (5 nm).(b) Bending in convex curvature directions produces tensile ($\varepsilon_{xx} > 0$) (c)concave bending produces compressive ($\varepsilon_{xx} < 0$) in-plane strain in the CFA film. The strain is applied parallel to the film plane along the bending direction.}
\label{fig:strain generation mechanism}
\end{figure*}

Al-rich Co–Fe–Al (CFA) thin films (25 nm) were deposited on flexible polyimide (PI) substrates using DC magnetron sputtering target under a base pressure below 4 × 10$^{-8}$ mbar. The film stack (the schematic is shown in Fig. ~\ref{fig:strain generation mechanism} (a)) consists of PI/Ta (15 nm)/CFA (25 nm)/Cu (5 nm), where the Ta seed layer improves surface smoothness, and the Cu capping layer prevents oxidation.The deposited CFA films are amorphous under the present deposition conditions. Details of the deposition conditions are explained in the Supplemental Material(SM) section S1 \cite{SM}. Uniaxial in-plane strain ($\varepsilon_{xx}$) in the range of 0–0.25\% was introduced via controlled bending of the flexible PI substrate. Here, $\varepsilon_{xx}$ denotes the Uniaxial in-plane strain ($\varepsilon_{xx}$) in the range of 0--0.25\% was introduced via controlled bending of the flexible PI substrate. Here, $\varepsilon_{xx}$ denotes
the strain parallel to the CFA film plane, with $\varepsilon_{xx} > 0$ corresponding to tensile strain and
$\varepsilon_{xx} < 0$ corresponding to compressive strain.
Both tensile and compressive strains were achieved using
concave and convex molds with different radii as shown in
Fig~\ref{fig:strain generation mechanism} (b) and (c), respectively. This in-plane strain configuration is consistent with established bending-strain approaches for thin films on flexible substrates \cite{li2023enhanced,Dai2013,tang2014magneto,Pandey_2020}. In addition, strain was also generated by depositing films on pre-bent substrates followed by flattening, enabling reproducible strain states. (Detailed strain generation method is shown in Fig. S1 in the SM \cite{SM})
 Surface topography and morphology were characterized using atomic force microscopy (AFM) and scanning electron microscopy (SEM) . The elemental composition of the CFA films was analyzed by energy-dispersive X-ray spectroscopy (EDXS) attached to the SEM, confirming Al enrichment. Magnetic properties were measured using superconducting quantum interference device (SQUID) magnetometry. Magnetization reversal and domain evolution under strain were investigated using longitudinal magneto-optical Kerr effect (MOKE) microscopy.

First-principles spin-polarized calculations were performed using Vienna Ab-initio Simulation Package (VASP) \cite{kresse1994ab,kresse1996efficient,kresse1999ultrasoft}. Exchange-correlation was treated within the generalized gradient approximation (GGA) using Perdew-Burke-Ernzerhof (PBE) functional \cite{perdew1996generalized,perdew2008restoring}. A plane-wave energy cutoff of 520 eV was adopted for all the calculations. Brillouin zone integrations were carried out using a Monkhorst-Pack k-point mesh of 11$\times$11$\times$11 for structural relaxations, while a denser k-mesh was employed for accurate total energy and electronic structure calculations. The total energy and interatomic forces were converged to 10$^{-6}$ and 0.001 eV/\AA. To account for on-site Coulomb interactions of the transition-metal (Co and Fe) $3d$ electrons, we adopted the DFT+$U$ approach using effective Hubbard parameters ($U_{\text{Co}}=3.5$~eV and $U_{\text{Fe}}=4.5$~eV) following the Dudarev scheme \cite{Dudarev}, chosen within the range reported in literature \cite{nawa2019exploring,capdevila2016performance,sasioglu2013ab}. It is seen that any change in the U parameters around these values does not impact the results significantly. To model the effect of Al enrichment (i.e ~52\% Al), a nominal Al$_2$CoFe configuration was constructed by modifying an inverse-Huesler framework to yield a 50\% Al occupancy on the Co sublattice.All strain-free reference structures were fully geometry-optimized (relaxing both unit cell dimensions and internal atomic coordinates) to establish their ground state, yielding an equilibrium lattice parameter of 5.69~\AA\ prior to the application of strain. More details are given in section S3 of SM.\cite{SM}

Surface morphology and topography characterized by SEM and AFM reveal uniform film growth on the flexible polyimide substrate, with no observable cracks, delamination, or strain-induced surface buckling up to the maximum applied strain of 0.25\%. The elemental composition of the deposited Co-Fe-Al films was first verified by energy-dispersive X-ray spectroscopy performed over multiple regions, confirming an intentionally Al-rich composition ($\approx$ 52 at.$\%$ Al). Detailed structural analysis is shown in Fig. S4 and S5 Of sections S4 of the SM \cite{SM}. 
\begin{figure}[!htbp]
\centering
\includegraphics[width=0.45\textwidth]{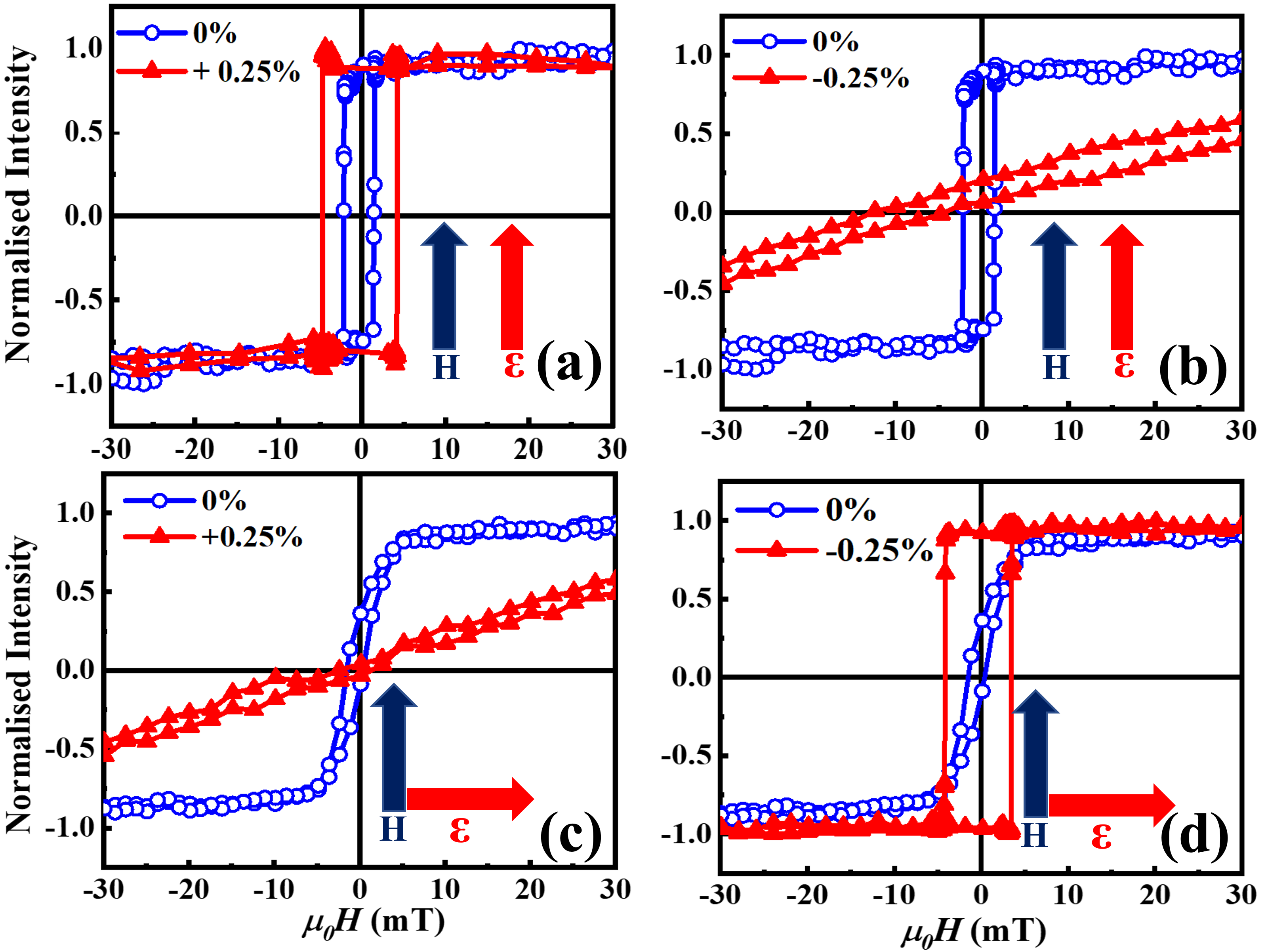}
\caption {Hysteresis loops of PI/CFA (25 nm) under applied strain: (a) tensile strain (+0.25 \%) with field parallel to the stress axis, (b) compressive strain (-0.25 \%) with field parallel to the stress axis, (c) tensile strain (+0.25 \%) with field perpendicular to the stress axis, and (d) compressive strain (-0.25 \%) with field perpendicular to the stress axis.}
\label{fig:hysteresis}
\end{figure}

\setlength{\textfloatsep}{5pt}
The magnetic response of the CFA films was investigated using SQUID-VSM and MOKE measurements under tensile and compressive strains. The applied strain was systematically varied from 0\% to 0.25\% by controlling the radius of curvature. Representative in-plane and out-of-plane hysteresis loops measured by SQUID confirm that the films retain dominant in-plane magnetic anisotropy, as evidenced by the higher saturation field and reduced remanence in the out-of-plane configuration (see Fig. S8 from SM \cite{SM}). MOKE hysteresis loops measured with the magnetic field applied parallel and perpendicular to the stress axis [Fig.~\ref{fig:hysteresis} (a–d)] reveal a clear strain-dependent evolution of the magnetic response. Under tensile strain, the parallel geometry exhibits enhanced coercivity and increased loop squareness, indicating stabilization of the easy axis along the strain direction [Fig.~\ref{fig:hysteresis} (a)]. In contrast, the perpendicular configuration shows reduced remanence and higher saturation fields, consistent with hard-axis behaviour [Fig.~\ref{fig:hysteresis} (c)]. Upon applying compressive strain, this trend reverses,i.e,  the parallel geometry evolves into a hard-axis-like response, while the perpendicular direction becomes magnetically favorable as shown in Fig.~\ref{fig:hysteresis} (b) and (c). Detailed strain-driven hysteresis loops evolution for 0.03\% and 0.12\% compressive and tensile strain are shown in the Fig. S6 of SM \cite{SM}.

\begin{figure}[!htbp]
\centering
\includegraphics[width=0.36\textwidth]{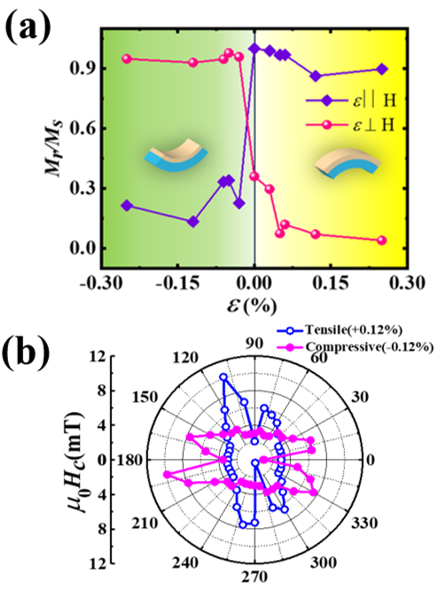}\caption {(a) Squareness ratio ($M_r/M_s$) vs strain along easy and hard axes. (b) Coercivity ($H_c$) under ±0.12\% tensile and compressive strain.}
\label{fig:strain_dependence}
\end{figure}

The anisotropy evolution is further quantified through the remanence ratio (\textit{M$_r$/M$_s$}) and angular dependence of coercivity. For \textit{H} ‖ $\mathcal{E}$, \textit {M$_r$/M$_s$} increases from 0.22  to 0.89, whereas in the perpendicular geometry it decreases to 0.03, confirming strong uniaxial anisotropy along the strain axis. Under compressive strain, the trend is reversed, with the perpendicular geometry reaching \textit{M$_r$/M$_s$} to 0.93 [Fig.~\ref{fig:strain_dependence} (a)]. The angular dependence of coercivity reveals a corresponding shift of the preferred magnetization reversal direction, evidencing a nearly 90$^\circ$ switching of the in-plane uniaxial anisotropy between tensile and compressive states as shown in Fig.~\ref{fig:strain_dependence} (b). Further, domain evolution exhibits the same trend, confirming strain-induced anisotropy switching [Fig. S7 in SM \cite{SM}]. This behavior can be understood within the magneto-elastic framework, where the anisotropy energy is given by magneto-elastic energy, $E_{ME}$ = 3/2($\lambda \sigma$ $\sin^2\theta$), governs the alignment of the easy axis relative to the applied stress, where $\sigma$ is the applied stress,$\lambda$ is the coefficient of magnetostriction,  and $\theta$ is the angle between the stress and magnetization direction \cite{cullity2011introduction,tang2014magneto}.

\begin{figure*}[!htbp]
\centering
\includegraphics[width=0.78\textwidth]{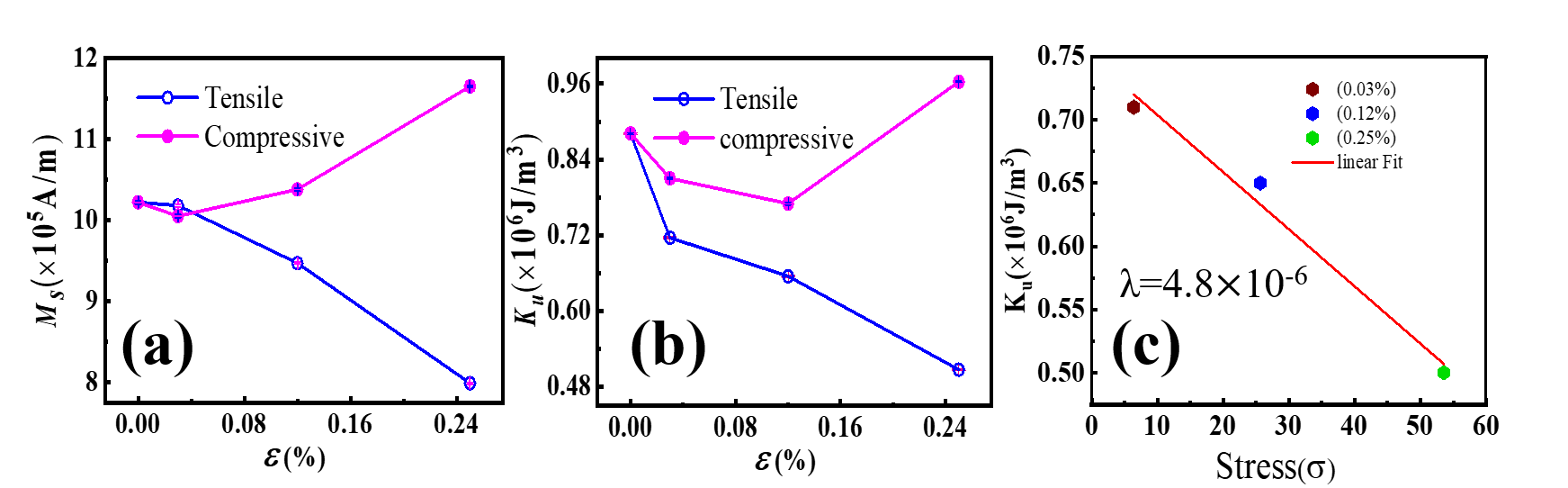}
\caption {The parameters were determined from the measured $M$–$H$ hysteresis data (a) Variation of saturation magnetization ($M_s$) with applied strain, showing an unconventional (inverted) strain response. (b) Strain dependence of the uniaxial anisotropy constant ($K_u$) under tensile and compressive strain. (c) Linear dependence of $K_u$ on stress, from which the magnetostriction coefficient ($\lambda$) of the CFA film is extracted.}
\label{fig:MagnetoAnisotropyConstant}
\end{figure*}

Remarkably, the saturation magnetization $M_s$ exhibits an unconventional strain dependence [Fig.~\ref{fig:MagnetoAnisotropyConstant} (a)]. Interestingly,
we observed that $M_s$ increases under compressive
strain but decreases under tensile strain, as shown
in Fig.~\ref{fig:MagnetoAnisotropyConstant} (a) which is opposite to trend typically reported. This inversion indicates a fundamentally 
modified magnetoelastic response in the Al-rich regime. It is known that when strain is applied, the resulting change in atomic spacing alters the hybridization of electronic orbitals, thereby modifying the band structure and spin-polarized states, which directly affects the exchange interactions governing the magnetic moment \mbox{\cite{jena2022strain,zhang2021strain,luo2018electronic,hirohata2020review,gao2018strain}}.
The strain-dependent anisotropy is quantified through the uniaxial anisotropy constant $K_u$, extracted from the hard-axis saturation field. As shown in Fig.~\ref{fig:MagnetoAnisotropyConstant} (b), $K_u$ decreases monotonically with tensile strain, whereas under compressive strain it exhibits a non-monotonic variation. Within the magnetoelastic framework, $K_u \propto \lambda \sigma$, yielding a positive magnetostriction coefficient for the film. The extracted value $\lambda$  = 4.8 $\times$ 10$^{-6}$, is consistent with previous reports and supports the strain-induced anisotropy switching discussed above, Fig.~\ref{fig:MagnetoAnisotropyConstant} (c) \cite{gueye2014bending,mahfouzi2020magnetoelastic}(see SM section S8 \cite{SM}). These results establish that, beyond anisotropy control, strain in Al-rich Co–Fe–Al films enables an unconventional inversion of magnetization, providing a key experimental signature of the modified electronic and magnetic interactions in this system.
\begin{figure}[!htbp]
\centering
\includegraphics[width=0.48\textwidth]{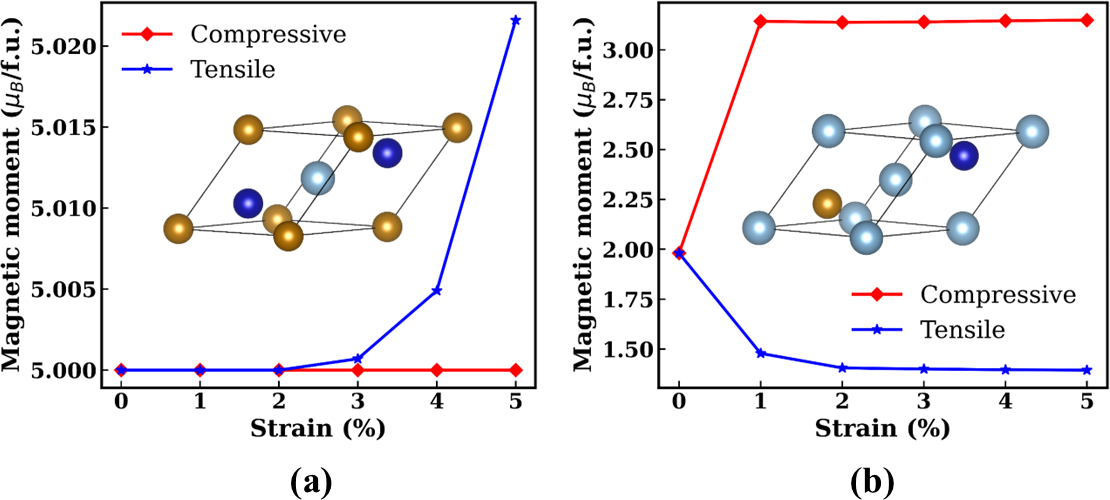}

\caption {Calculated total magnetic moment as a function of uniaxial strain for (a) Co$_2$FeAl and (b) Al$_2$CoFe. Stoichiometric Co$_2$FeAl displays weak, conventional strain dependence, whereas the Al-rich configuration exhibits a qualitatively reversed response, reflecting the strain-sensitive magnetic instability of the inverse-Heusler–like environment.}
\label{fig:magmom}
\end{figure}
\setlength{\textfloatsep}{5pt}
To elucidate the microscopic origin of the unconventional strain response observed experimentally in Al-rich Fe–Co–Al films, we performed first-principles calculations focusing on the role of Al enrichment and the resulting local chemical environments. While the sputtered films deposited on flexible substrates lack long-range chemical order, their magnetic response is governed primarily by the local coordination of  Fe and Co atoms. Ordered reference structures are therefore employed as representative models to capture the essential local bonding motifs and hybridization effects introduced by excess Al.
Stoichiometric Co$_2$FeAl shows a robust ferromagnetic state driven by strong  $d-d$ exchange interactions due to the dominant transition-metal environment around Fe and Co sublattices. Under the application of uniaxial strain Fig.\ref{fig:magmom} (a), we observe that the magnetic moments vary weakly and monotonically, preserving ferromagnetic order.  The calculated site-projected moments (Table 1) confirm robust ferromagnetic alignment, with parallel Co and Fe moments and an integer total moment ($\approx 5~\mu_{\mathrm{B}}$)consistent with Slater-Pauling expectations.  
To model the experimentally realized Al-rich composition ( $\approx$ 52 at.\% Al), we consider a nominal Al$_2$CoFe configuration derived from inverse-Heusler ordering, with an optimized lattice parameter of 5.69 \AA. The structure is both dynamically and thermodynamically stable, as the phonon dispersion calculations reveal no imaginary modes, and the computed formation energy is negative (see  Fig. S3 in SM \cite{SM}).  In contrast to L2$_1$ Co$_2$FeAl phase, the Al-enriched configuration exhibits a metallic ground state with a pronounced mixing between Al-sp states and transition-metal d states. The calculated site-projected magnetic moments reveal antiparallel Co-Fe coupling, with Co carrying a negative moment ($\sim$ - 0.708 $\mu_B$) and Fe a positive moment ($\sim$  1.887 $\mu_B$), resulting in a substantially reduced total moment per formula unit, as shown in Table \ref{tab:moments} in SM \cite{SM}. The absence of an integer magnetic moment indicates itinerant Stoner-type magnetism rather than rigid Slater–Pauling behavior. The structural origin of this modification lies in the redistribution of local coordination.  In particular, ‘Al’ becomes the dominant nearest neighbour, and Co-Fe separation increases relative to the L2$_1$ Co$_2$FeAl phase (Table S2, in SM \cite{SM}). This enhances $d–sp$ hybridization while diminishing the direct $d-d$ exchange interactions, thereby destabilizing the high-spin ferromagnetic configuration.
 Although these models do not represent the exact amorphous structure of the experimental films, they capture the dominant local chemical rearrangements induced by Al enrichment and provide a meaningful framework for analyzing strain-dependent magnetic behavior.  

\begin{table}[t]
\centering
\renewcommand{\arraystretch}{1.15}
\begin{tabular}{c|c|c|c|c|c}
\hline
 &  & \multicolumn{2}{c|}{Co$_2$FeAl} & \multicolumn{2}{c}{Al$_2$CoFe} \\
\hline
Regime & Strain & $\mu_B(\mathrm{Co})$ & $\mu_B(\mathrm{Fe})$ & $\mu_B(\mathrm{Co})$ & $\mu_B(\mathrm{Fe})$ \\
\hline
\multirow{5}{*}{\rotatebox{90}{Compressive}} 
& $-5\%$ & 1.158 & 2.930 & 0.861  & 2.665 \\
& $-4\%$ & 1.160 & 2.937 & 0.876  & 2.685 \\
& $-3\%$ & 1.163 & 2.943 & 0.889  & 2.709 \\
& $-2\%$ & 1.165 & 2.950 & 0.898  & 2.718 \\
& $-1\%$ & 1.167 & 2.957 & 0.908  & 2.729 \\
\hline
 & \textbf{0\%}  & 1.169 & 2.964 & -0.708 & 1.887 \\
\hline
\multirow{5}{*}{\rotatebox{90}{Tensile}} 
& $1\%$  & 1.171 & 2.970 & -0.900 & 2.492 \\
& $2\%$  & 1.174 & 2.977 & -0.601 & 2.486 \\
& $3\%$  & 1.177 & 2.983 & -0.785 & 2.531 \\
& $4\%$  & 1.182 & 2.990 & -0.911 & 2.565 \\
& $5\%$  & 1.188 & 2.996 & -1.157 & 2.565 \\
\hline
\end{tabular}
\caption{DFT-calculated sublattice-resolved magnetic moments (in $\mu_B$) of Co and Fe as a function of epitaxial strain for stoichiometric Co$_2$FeAl (L2$_1$ Heusler) and Al-rich Al$_2$CoFe (inverse-Heusler-like). The leftmost column denotes the strain regime (compressive/tensile), and the zero-strain reference point is highlighted in bold. Notably, Al$_2$CoFe exhibits a strong strain-induced sublattice compensation, with the Co moment changing sign between compressive and tensile strain while the Fe moment remains positive, leading to a reduced net moment under tensile strain.}
\label{tab:moments}
\end{table}

A striking contrast emerges under strain in the Al-rich configuration. Unlike stoichiometric Co$_2$FeAl, the Al$_2$CoFe model displays an inverted strain response: compressive strain enhances the total magnetic moment, while tensile strain suppresses it [Fig. \ref{fig:magmom} (b)]. This behavior is consistent with the experiment [Fig.~\ref{fig:MagnetoAnisotropyConstant} (a)].  The origin of this inversion can be traced to the instability of the antiparallel Co-Fe alignment. Under compressive strain, the Co moment undergoes a spin-flip  (sign reversal, Table \ref{tab:moments}) and aligns parallel to the Fe moment, resulting in a pronounced increase in the increase in the total magnetization. Conversely, tensile strain stabilizes the antiparallel alignment and suppresses the net moment. Hence, the strain-driven reconfiguration of the Co moment provides a microscopic explanation for the experimentally observed enhancement of magnetization under compression.

The electronic origin is clarified by the projected density of states (PDOS) in the SM (Fig. S9) \cite{SM}. In the ideal Co$_2$FeAl (L2$_1$) phase, the exchange splitting of the transition-metal $d$ states is significantly larger than the crystal-field splitting, yielding a stable high-spin configuration. The DOS near the Fermi level remains relatively insensitive to applied strain as shown in Fig. S9 (a-c) \cite{SM}.  In contrast, the Al-rich configuration exhibits sharp $d$-derived features near the Fermi level, indicating that a system is close to magnetic instability. Compressive strain broadens these features and shifts spectral weight away from the Fermi level, stabilizing the high-spin ferromagnetic state. However, tensile strain sharpens the DOS near E$_F$, enhancing the relative influence of crystal-field effects over exchange and promoting moment suppression. Such strain-induced DOS reshaping provides a direct electronic mechanism for strong magnetoelastic response observed experimentally.

\par To rationalize the strain dependence of the magnetic moment obtained from DFT, it is useful to distinguish between the nearly conventional response of stoichiometric \(\mathrm{Co_2FeAl}\) and the anomalous behavior of the Al-rich inverse-Heusler-like configuration \(\mathrm{Al_2CoFe}\). In stoichiometric \(\mathrm{Co_2FeAl}\), the Co and Fe moments remain positive and evolve only weakly with strain, consistent with a robust ferromagnetic state. By contrast, in \(\mathrm{Al_2CoFe}\) the strain dependence is qualitatively different: the Fe moment remains positive throughout the considered strain window, whereas the Co moment is much more strain sensitive. In particular, the Co moment is positive in the compressive regime, becomes negative near zero strain, and remains negative over the tensile regime. The reduction of the net moment under tensile strain therefore does not arise from a uniform suppression of magnetism on all sites, but rather from enhanced compensation between inequivalent Co and Fe contributions.

This behavior cannot be captured naturally within a one-order-parameter Landau theory, which only describes the strengthening or weakening of a single collective magnetic amplitude. A minimal phenomenological description instead requires two coarse-grained sublattice magnetizations, \(m_{\mathrm{Co}}\) and \(m_{\mathrm{Fe}}\), associated with the Co- and Fe-dominated sectors. We therefore introduce the Landau free energy
\begin{equation}
\begin{aligned}
F(m_{\mathrm{Co}},m_{\mathrm{Fe}};\varepsilon)
=&
\frac{1}{2}a_{\mathrm{Co}}(\varepsilon)m_{\mathrm{Co}}^2
+\frac{1}{4}b_{\mathrm{Co}}m_{\mathrm{Co}}^4
+\frac{1}{2}a_{\mathrm{Fe}}(\varepsilon)m_{\mathrm{Fe}}^2\\
&+\frac{1}{4}b_{\mathrm{Fe}}m_{\mathrm{Fe}}^4
+J(\varepsilon)m_{\mathrm{Co}}m_{\mathrm{Fe}}.
\label{eq:landau_two_sublattice}
\end{aligned}
\end{equation}
Here \(\varepsilon\) denotes the externally applied strain, \(a_{\mathrm{Co}}\) and \(a_{\mathrm{Fe}}\) are quadratic coefficients controlling the local tendency toward moment formation on the two sublattices, \(b_{\mathrm{Co}},b_{\mathrm{Fe}}>0\) ensure stability, and \(J(\varepsilon)\) is an effective inter-sublattice exchange coupling. In the present sign convention, \(J<0\) favors parallel alignment whereas \(J>0\) favors antiparallel alignment. The strain dependence of these coefficients encodes, at a coarse-grained level, the effect of lattice distortion on exchange splitting, crystal-field energies, and local hybridization. To make the strain dependence explicit at lowest order, these coefficients may be written as
\begin{equation}
\begin{split}
a_{\mathrm{Co}}(\varepsilon)&=a_{\mathrm{Co}}^{(0)}+\alpha_{\mathrm{Co}}\varepsilon,\\
a_{\mathrm{Fe}}(\varepsilon)&=a_{\mathrm{Fe}}^{(0)}+\alpha_{\mathrm{Fe}}\varepsilon,\\
J(\varepsilon)&=J_0+J_1\varepsilon,
\end{split}
\label{eq:landau_linear_coeff}
\end{equation}
where the coefficients are introduced only phenomenologically and are not fitted to the DFT data.

The equilibrium moments follow from minimizing the free energy:
\begin{equation}
a_{\mathrm{Co}}(\varepsilon)m_{\mathrm{Co}}+b_{\mathrm{Co}}m_{\mathrm{Co}}^3+J(\varepsilon)m_{\mathrm{Fe}}=0,
\label{eq:mco_stationary}
\end{equation}
\begin{equation}
a_{\mathrm{Fe}}(\varepsilon)m_{\mathrm{Fe}}+b_{\mathrm{Fe}}m_{\mathrm{Fe}}^3+J(\varepsilon)m_{\mathrm{Co}}=0.
\label{eq:mfe_stationary}
\end{equation}
The distinction between the regular and Al-rich structures becomes explicit by differentiating these equilibrium conditions with respect to strain. This gives
\begin{equation}
\mathbf{H}_{m}
\begin{pmatrix}
\mathrm{d}m_{\mathrm{Co}}/\mathrm{d}\varepsilon\\[1mm]
\mathrm{d}m_{\mathrm{Fe}}/\mathrm{d}\varepsilon
\end{pmatrix}
=-
\begin{pmatrix}
a'_{\mathrm{Co}}m_{\mathrm{Co}}+J'm_{\mathrm{Fe}}\\
a'_{\mathrm{Fe}}m_{\mathrm{Fe}}+J'm_{\mathrm{Co}}
\end{pmatrix},
\label{eq:strain_response}
\end{equation}
where primes denote derivatives with respect to strain and
\begin{equation}
\mathbf{H}_{m}=
\begin{pmatrix}
a_{\mathrm{Co}}+3b_{\mathrm{Co}}m_{\mathrm{Co}}^2 & J\\
J & a_{\mathrm{Fe}}+3b_{\mathrm{Fe}}m_{\mathrm{Fe}}^2
\end{pmatrix}
\label{eq:magnetic_stiffness}
\end{equation}
is the magnetic-stiffness matrix, i.e., the Hessian of the free energy evaluated at the equilibrium state. Equation~\ref{eq:strain_response} shows that the magnetic response to strain depends on both the explicit strain dependence of the electronic interactions and the curvature of the magnetic free-energy minimum.

This provides the qualitative distinction between the two structures that is central to the present work. In the regular L2$_1$-Co$_2$FeAl phase, the large exchange splitting and weak strain dependence of the electronic states near $E=E_F$ indicate a comparatively deep and stiff ferromagnetic minimum. Within the Landau picture, this corresponds to a magnetic-stiffness matrix $\mathbf H_m$ whose inverse does not strongly amplify the strain-induced perturbation. Consequently, the changes in both Co and Fe moments remain small, in agreement with Table~\ref{tab:moments} and Fig.~\ref{fig:magmom} (a). In the Al-rich inverse-Heusler-like configuration, by contrast, Al-induced \(sp\)--\(d\) hybridization weakens the rigid transition-metal ferromagnetic state and brings a Co-dominated compensation mode close to magnetic instability. The corresponding curvature of the free-energy landscape is smaller, so that \(\mathbf{H}_{m}^{-1}\) amplifies the effect of strain on the sublattice moments. This is consistent with the sharp strain dependence of the near-\(E_F\) DOS and with the calculated sign reversal of the Co moment. Thus, the two systems occupy qualitatively different regions of the same Landau parameter space: stoichiometric Co$_2$FeAl lies well inside a robust ferromagnetic basin, whereas Al$_2$CoFe lies close to a soft sublattice-compensation instability.

For the Al-rich configuration, where the soft magnetic mode is predominantly associated with the Co sector and the Fe moment remains comparatively robust, the mechanism can also be seen directly from the first equilibrium equation. When \(m_{\mathrm{Co}}\) remains moderate,
\begin{equation}
m_{\mathrm{Co}}\approx -\frac{J(\varepsilon)}{a_{\mathrm{Co}}(\varepsilon)}\,m_{\mathrm{Fe}}.
\label{eq:mco_approx}
\end{equation}
Tensile strain can therefore reduce the net moment either by increasing the effective Co stiffness \(a_{\mathrm{Co}}(\varepsilon)\), by enhancing the tendency toward antiparallel alignment through \(J(\varepsilon)\), or through a combination of both. The resulting picture is one of strain-sensitive sublattice compensation rather than a uniform collapse of ferromagnetism.

The present Landau construction is intended as a minimal interpretive framework rather than a quantitative fit. The DFT sublattice moments in Al$_2$CoFe show a relatively sharp Co-moment sign reversal between the compressive regime and the vicinity of zero strain, together with some non-monotonic evolution on the tensile side (Table~\ref{tab:moments}). Such details reflect microscopic electronic-structure effects beyond this lowest-order phenomenology. Introducing the net and compensating combinations \(M=m_{\mathrm{Co}}+m_{\mathrm{Fe}}\) and \(L=m_{\mathrm{Fe}}-m_{\mathrm{Co}}\), tensile strain can be viewed as transferring weight toward the compensating channel \(L\), whereas compression favors the net-moment channel \(M\). Additional discussion of this interpretation and its limitations is retained in the updated SM SI9 \cite{SM}.

In summary,  we demonstrate a strong and unconventional strain-controlled magnetic response in Al-rich Fe–Co–Al thin films on flexible polyimide substrates. Mechanical bending enables reversible tuning of magnetic anisotropy, coercivity, and saturation magnetization. The Al-rich Co-Fe-Al films exhibit an inverted strain dependence, where compressive strain enhances the magnetization while tensile strain suppresses it. Strain also induced a clear rotation of the uniaxial easy axis, confirming strong magnetoelastic coupling.
First-principles calculations reveal that this behavior arises from the Al-rich inverse-Heusler-like local environment, in which strain modifies the crystal-field splitting and $d–d$ hybridization, leading to competing spin states. These results establish Al-rich Fe–Co–Al as a promising platform for strain-programmable flexible spintronic and magnetic sensing applications.
\vspace{-8pt}
\section{Acknowledgments}
We thank the Department of Atomic Energy (No. 0803/2/2020/NISER/R\&D-II/8149), Department of Science and Technology, Science and Engineering Research Board (Grant No.CRG/2021/001245) for providing financial support. This work was also supported by KIST (Korea Institute of Science and Technology) Institutional Program: Global Knowledge Platform Research Project (26Z0100).

\bibliography{References} 

@article{phonopy-phono3py-JPCM,
  author = {A. Togo and L. Chaput and T. Tadano and I. Tanaka},
  title = {Implementation strategies in phonopy and phono3py},
  journal = {J. Phys.: Condens. Matter},
  volume = {35},
  pages = {353001},
  year = {2023},
  doi = {10.1088/1361-648X/acd831}
}

@article{phonopy-phono3py-JPSJ,
    author = {A. Togo},
  title = {First-principles Phonon Calculations with Phonopy and Phono3py},
  journal = {J. Phys. Soc. Jpn.},
  volume = {92},
  pages = {012001},
  year = {2023},
  doi = {10.7566/JPSJ.92.012001}
}

@article{corzo2020flexible,
  author = {Corzo, Daniel and Tostado-Bl{\'a}zquez, Guillermo and Baran, Derya},
  title = {Flexible Electronics: Status, Challenges and Opportunities},
  journal = {Front. in Electron.},
  volume = {1},
  pages = {594003},
  year = {2020},
  doi = {10.3389/felec.2020.594003}
}

@article{hwang2012flexible,
  author = {Sun Kak Hwang and Insung Bae and Richard Hahnkee Kim and Cheolmin Park},
  title = {Flexible Non-Volatile Ferroelectric Polymer Memory with Gate-Controlled Multilevel Operation},
  journal = {Adv. Mater.},
  volume = {24},
  pages = {5910--5914},
  year = {2012},
  doi = {10.1002/adma.201201831}
}

@article{han2013layer,
  author = {Su-Ting Han and Ye Zhou and V. A. L. Roy},
  title = {Towards the Development of Flexible Non-Volatile Memories},
  journal = {Adv. Mater.},
  volume = {25},
  pages = {5425--5449},
  year = {2013},
  doi = {10.1002/adma.201301361}
}

@article{schafer2009suppression,
title = {Suppression of de-wetting of copper coatings on carbon substrates by metal (Cr, Mo, Ti) doped boron interlayers},
author = {D. Schäfer and J. Hell and C. Eisenmenger-Sittner and E. Neubauer and H. Hutter and N. Kornfeind},
journal = {Vacuum},
volume = {84},
number = {1},
pages = {202-204},
year = {2009},
doi = {https://doi.org/10.1016/j.vacuum.2009.04.017},
url = {https://www.sciencedirect.com/science/article/pii/S0042207X09002012},
}

@article{avc2008influence,
   author = {V. {\'A}{\v{c}} and B. Anwarzai and S. Luby and E. Majkova},
  title = {Influence of mechanical strain on magnetic characteristics of spin valves},
  journal = {J. Phys.: Conf. Ser.},
  volume = {100},
  pages = {082025},
  year = {2008},
  doi = {10.1088/1742-6596/100/8/082025}
}

@article{sheng2018flexible,
  author = {P. Sheng and B. Wang and R. Li},
  title = {Flexible magnetic thin films and devices},
  journal = {J. Semicond.},
  volume = {39},
  pages = {011006},
  year = {2018},
  doi = {10.1088/1674-4926/39/1/011006}
}

@article{yang2024advances,
  author = {H. Yang and S. Li and Y. Wu and X. Bao and Z. Xiang and Y. Xie and L. Pan and J. Chen and Y. Liu and R.-W. Li},
  title = {Advances in flexible magnetosensitive materials and devices for wearable electronics},
  journal = {Adv. Mater.},
  volume = {36},
  pages = {2311996},
  year = {2024},
  doi = {10.1002/adma.202311996}
}

@article{chen2021mechanically,
    author = {Chen, Xia and Mi, Wenbo},
    title = {Mechanically tunable magnetic and electronic transport properties of flexible magnetic films and their heterostructures for spintronics},
    journal = {J. Mater. Chem. C},
    volume = {9},
    number = {30},
    pages = {9400-9430},
    year = {2021},
    month = {08},
    doi = {10.1039/d1tc01989a},
    url = {https://doi.org/10.1039/d1tc01989a},
}

@article{zhao2018low,
  author = {S. Zhao and Z. Zhou and C. Li and B. Peng and Z. Hu and M. Liu},
  title = {Low-voltage control of (Co/Pt)$_x$ perpendicular magnetic anisotropy heterostructure for flexible spintronics},
  journal = {ACS Nano},
  volume = {12},
  pages = {7167--7173},
  year = {2018},
  doi = {10.1021/acsnano.8b03153}
}

@article{emori2017coexistence,
  author = {S. Emori and B. A. Gray and H.-M. Jeon and J. Peoples and M. Schmitt and K. Mahalingam and M. Hill and M. E. McConney and M. T. Gray and U. S. Alaan and A. C. Bornstein and P. Shafer and A. T. N'Diaye and E. Arenholz and G. Haugstad and K.-Y. Meng and F. Yang and D. Li and S. Mahat and D. G. Cahill and P. Dhagat and A. Jander and N. X. Sun and Y. Suzuki and B. M. Howe},
  title = {Coexistence of Low Damping and Strong Magnetoelastic Coupling in Epitaxial Spinel Ferrite Thin Films},
  journal = {Adv. Mater.},
  volume = {29},
  pages = {1701130},
  year = {2017},
  doi = {10.1002/adma.201701130}
}

@article{finizio2014magnetic,
  author = {S. Finizio and M. Foerster and M. Buzzi and B. Kr{\"u}ger and M. Jourdan and C. A. F. Vaz and J. Hockel and T. Miyawaki and A. Tkach and S. Valencia and F. Kronast and G. P. Carman and F. Nolting and M. Klaui},
  title = {Magnetic anisotropy engineering in thin film Ni nanostructures by magnetoelastic coupling},
  journal = {Phys. Rev. Applied},
  volume = {1},
  pages = {021001},
  year = {2014},
  doi = {10.1103/PhysRevApplied.1.021001}
}

@article{begue2021strain,
    author = {Begué, Adrián and Ciria, Miguel},
    title = {Strain-Mediated
Giant Magnetoelectric Coupling in
a Crystalline Multiferroic Heterostructure},
    journal = {ACS Applied Materials \& Interfaces},
    volume = {13},
    number = {5},
    pages = {6778-6784},
    year = {2021},
    month = {02},
    doi = {10.1021/acsami.0c18777},
    url = {https://doi.org/10.1021/acsami.0c18777},
}

@article{heuver2015strain,
  author = {J. A. Heuver and A. Scaramucci and Y. Blickenstorfer and S. Matzen and N. A. Spaldin and C. Ederer and B. Noheda},
  title = {Strain-induced magnetic anisotropy in epitaxial thin films of the spinel CoCr$_2$O$_4$},
  journal = {Phys. Rev. B},
  volume = {92},
  pages = {214429},
  year = {2015},
  doi = {10.1103/PhysRevB.92.214429}
}

@article{Liu_2013,
doi = {10.1088/1674-1056/22/12/127502},
url = {https://doi.org/10.1088/1674-1056/22/12/127502},
year = {2013},
month = {dec},
publisher = {},
volume = {22},
number = {12},
pages = {127502},
author = {Liu Yi-Wei,Zhan Qing-Feng,Li Run-Wei},
title = {Fabrication, properties, and applications of flexible magnetic films},
journal = {Chin. Phys. B}
}

@article{wen2019spin,
  author = {Z. Wen and Z. Qiu and S. T{\"o}lle and C. Gorini and T. Seki and D. Hou and T. Kubota and U. Eckern and E. Saitoh and K. Takanashi},
  title = {Spin-charge conversion in NiMnSb Heusler alloy films},
  journal = {Sci. Adv.},
  volume = {5},
  pages = {eaaw9337},
  year = {2019},
  doi = {10.1126/sciadv.aaw9337}
}

@article{wurmehl2006investigation,
    author = {Wurmehl, Sabine and Fecher, Gerhard H. and Kandpal, Hem Chandra and Ksenofontov, Vadim and Felser, Claudia and Lin, Hong Ji},
    title = {Investigation of Co$_2$FeSi: The Heusler compound with highest Curie temperature and magnetic moment},
    journal = {Appl. Phys. Lett.},
    volume = {88},
    number = {3},
    pages = {032503},
    year = {2006},
    month = {01},
    doi = {10.1063/1.2166205},
    url = {https://doi.org/10.1063/1.2166205},
}

@article{chumak2021magnetoelastic,
  author = {O. M. Chumak and A. Pacewicz and A. Lynnyk and B. Salski and T. Yamamoto and T. Seki and J. Z. Domagala and H. G{\l}owi{\'n}ski and K. Takanashi and L. T. Baczewski and H. Szymczak and A. Nabialek},
  title = {Magnetoelastic interactions and magnetic damping in Co$_2$Fe$_{0.4}$Mn$_{0.6}$Si and Co$_2$FeGa$_{0.5}$Ge$_{0.5}$ Heusler alloys thin films for spintronic applications},
  journal = {Sci. Rep.},
  volume = {11},
  pages = {7608},
  year = {2021},
  doi = {10.1038/s41598-021-87205-y}
}

@article{zhang2018direct,
  author = {X. Zhang and H. Xu and B. Lai and Q. Lu and X. Lu and Y. Chen and W. Niu and C. Gu and W. Liu and X. Wang and C. Liu and Y. Nie and L. He and Y. Xu},
  title = {Direct observation of high spin polarization in Co$_2$FeAl thin films},
  journal = {Sci. Rep.},
  volume = {8},
  pages = {8074},
  year = {2018},
  doi = {10.1038/s41598-018-26285-9}
}

@article{titov2018microstructure,
    author = {Titov, A. and Jiraskova, Y. and Zivotsky, O. and Bursik, J. and Janickovic, D.},
    title = {Microstructure and magnetism of Co$_2$FeAl Heusler alloy prepared by arc and induction melting compared with planar flow casting},
    journal = {AIP Adv.},
    volume = {8},
    number = {4},
    pages = {047206},
    year = {2017},
    month = {10},
    issn = {2158-3226},
    doi = {10.1063/1.4993698},
    url = {https://doi.org/10.1063/1.4993698},
}

@article{kudryavtsev2007evolution,
  author = {Y. V. Kudryavtsev and V. A. Oksenenko and Y. P. Lee and Y. H. Hyun and J. B. Kim and J. S. Park and S. Y. Park and J. Dubowik},
  title = {Evolution of the magnetic properties of Co$_2$MnGa Heusler alloy films: From amorphous to ordered films},
  journal = {Phys. Rev. B},
  volume = {76},
  pages = {024430},
  year = {2007},
  doi = {10.1103/PhysRevB.76.024430}
}

@article{ahmad2020competition,
  author = {A. Ahmad and A. K. Das and S. K. Srivastava},
  title = {Competition of L2$_1$ and XA ordering in Fe$_2$CoAl Heusler alloy: A first-principles study},
  journal = {Eur. Phys. J. B},
  volume = {93},
  pages = {96},
  year = {2020},
  doi = {10.1140/epjb/e2020-100626-4}
}

@article{rai2020pressure,
    author = {Rai, D. P. and Lalrinkima and Lalhriatzuala and Fomin, L. A. and Malikov, I. V. and Sayede, Adlane and Ghimire, Madhav Prasad and Thapa, R. K. and Zadeng, Lalthakimi},
    title = {Pressure dependent half-metallic ferromagnetism in inverse Heusler alloy Fe2CoAl: a DFT+U calculations},
    journal = {RSC Adv.},
    volume = {10},
    number = {73},
    pages = {44633-44640},
    year = {2020},
    month = {12},
    issn = {2046-2069},
    doi = {10.1039/d0ra07543d},
    url = {https://doi.org/10.1039/d0ra07543d},
}

@article{siakeng2018electronic,
    author = {Siakeng, Lalrinkima and Mikhailov, Gennady M. and Rai, D. P.},
    title = {Electronic, elastic and X-ray spectroscopic properties of direct and inverse full Heusler compounds Co2FeAl and Fe2CoAl, promising materials for spintronic applications: a DFT+U approach},
    journal = {J. Mater. Chem. C},
    volume = {6},
    number = {38},
    pages = {10341-10349},
    year = {2018},
    month = {10},
    doi = {10.1039/c8tc02530d},   
}

@article{ahmad2020phase,
  author = {A. Ahmad and S. K. Srivastava and A. K. Das},
  title = {Phase stability and the effect of lattice distortions on electronic properties and half-metallic ferromagnetism of Co$_2$FeAl Heusler alloy: An ab initio study},
  journal = {J. Phys.: Condens. Matter},
  volume = {32},
  pages = {415606},
  year = {2020},
  doi = {10.1088/1361-648X/ab9f4f}
}

@misc{SM,
title={Supplemental Material},
}

@article{li2023enhanced,
    author = {Li, Mengchao and Yang, Huali and Xie, Yali and Huang, Kai and Pan, Lili and Tang, Wei and Bao, Xilai and Yang, Yumeng and Sun, Jie and Wang, Xinming and Che, Shenglei and Li, Run-Wei},
    title = {Enhanced Stress Stability in Flexible Co/Pt Multilayers
with Strong Perpendicular Magnetic Anisotropy},
    journal = {Nano Lett.},
    volume = {23},
    number = {17},
    pages = {8073-8080},
    year = {2023},
    month = {09},
    issn = {1530-6984},
    doi = {10.1021/acs.nanolett.3c02047},
    url = {https://doi.org/10.1021/acs.nanolett.3c02047},
    
}

@article{Dai2013,
  author = {G. Dai and Q. Zhan and H. Yang and Y. Liu and X. Zhang and Z. Zuo and B. Chen and R.-W. Li},
  title = {Controllable strain-induced uniaxial anisotropy of Fe$_{81}$Ga$_{19}$ films deposited on flexible bowed-substrates},
  journal = {J. Appl. Phys.},
  volume = {114},
  pages = {173913},
  year = {2013},
  doi = {10.1063/1.4829670}
}

@article{tang2014magneto,
  author = {Z. Tang and B. Wang and H. Yang and X. Xu and Y. Liu and D. Sun and L. Xia and Q. Zhan and B. Chen and M. Tang and Y. Zhou and J. Wang and R.-W. Li},
  title = {Magneto-mechanical coupling effect in amorphous Co$_{40}$Fe$_{40}$B$_{20}$ films grown on flexible substrates},
  journal = {Appl. Phys. Lett.},
  volume = {105},
  pages = {103504},
  year = {2014},
  doi = {10.1063/1.4895628}
}

@article{Pandey_2020,
doi = {10.1088/2632-959X/ab90cb},
url = {https://doi.org/10.1088/2632-959X/ab90cb},
year = {2020},
month = {may},
publisher = {IOP Publishing},
volume = {1},
number = {1},
pages = {010037},
author = {Pandey, Esita and Singh, Braj Bhusan and Sharangi, Purbasha and Bedanta, Subhankar},
title = {Strain engineered domain structure and their relaxation in perpendicularly magnetized Co/Pt deposited on flexible polyimide},
journal = {Nano Express},
}

@article{kresse1994ab,
  title = {Ab initio molecular-dynamics simulation of the liquid-metal--amorphous-semiconductor transition in germanium},
  author = {Kresse, G. and Hafner, J.},
  journal = {Phys. Rev. B},
  volume = {49},
  issue = {20},
  pages = {14251--14269},
  numpages = {0},
  year = {1994},
  month = {May},
  publisher = {American Physical Society},
  doi = {10.1103/PhysRevB.49.14251},
}

@article{kresse1996efficient,
  title = {Efficient iterative schemes for ab initio total-energy calculations using a plane-wave basis set},
  author = {Kresse, G. and Furthm\"uller, J.},
  journal = {Phys. Rev. B},
  volume = {54},
  issue = {16},
  pages = {11169--11186},
  numpages = {0},
  year = {1996},
  month = {Oct},
  publisher = {American Physical Society},
  doi = {10.1103/PhysRevB.54.11169},
}

@article{kresse1999ultrasoft,
  title = {From ultrasoft pseudopotentials to the projector augmented-wave method},
  author = {Kresse, G. and Joubert, D.},
  journal = {Phys. Rev. B},
  volume = {59},
  issue = {3},
  pages = {1758--1775},
  numpages = {0},
  year = {1999},
  month = {Jan},
  publisher = {American Physical Society},
  doi = {10.1103/PhysRevB.59.1758},

}

@article{perdew1996generalized,
  title = {Generalized Gradient Approximation Made Simple},
  author = {Perdew, John P. and Burke, Kieron and Ernzerhof, Matthias},
  journal = {Phys. Rev. Lett.},
  volume = {77},
  issue = {18},
  pages = {3865--3868},
  numpages = {0},
  year = {1996},
  month = {Oct},
  publisher = {American Physical Society},
  doi = {10.1103/PhysRevLett.77.3865},
}

@article{perdew2008restoring,
  title = {Restoring the Density-Gradient Expansion for Exchange in Solids and Surfaces},
  author = {Perdew, John P. and Ruzsinszky, Adrienn and Csonka, G\'abor I. and Vydrov, Oleg A. and Scuseria, Gustavo E. and Constantin, Lucian A. and Zhou, Xiaolan and Burke, Kieron},
  journal = {Phys. Rev. Lett.},
  volume = {100},
  issue = {13},
  pages = {136406},
  numpages = {4},
  year = {2008},
  month = {Apr},
  publisher = {American Physical Society},
  doi = {10.1103/PhysRevLett.100.136406},
}

@article{Dudarev,
  title = {Electron-energy-loss spectra and the structural stability of nickel oxide:  An LSDA+U study},
  author = {Dudarev, S. L. and Botton, G. A. and Savrasov, S. Y. and Humphreys, C. J. and Sutton, A. P.},
  journal = {Phys. Rev. B},
  volume = {57},
  issue = {3},
  pages = {1505--1509},
  numpages = {0},
  year = {1998},
  month = {Jan},
  publisher = {American Physical Society},
  doi = {10.1103/PhysRevB.57.1505},
  url = {https://link.aps.org/doi/10.1103/PhysRevB.57.1505}
}

@article{nawa2019exploring,
    author = {Nawa, Kenji and Miura, Yoshio},
    title = {Exploring half-metallic Co-based full Heusler alloys using a DFT+U method combined with linear response approach},
    journal = {RSC Adv.},
    volume = {9},
    number = {52},
    pages = {30462-30478},
    year = {2019},
    month = {09},
    doi = {10.1039/c9ra05212g},
    url = {https://doi.org/10.1039/c9ra05212g},
}

@article{capdevila2016performance,
    author = {Capdevila-Cortada, Marçal and Łodziana, Zbigniew and López, Núria},
    title = {Performance of DFT+U Approaches in
the Study of Catalytic Materials},
    journal = {ACS Catal.},
    volume = {6},
    number = {12},
    pages = {8370-8379},
    year = {2016},
    month = {12},
    issn = {2155-5435},
    doi = {10.1021/acscatal.6b01907},
    url = {https://doi.org/10.1021/acscatal.6b01907},
}

@article{sasioglu2013ab,
  title = {Ab initio calculation of the effective on-site Coulomb interaction parameters for half-metallic magnets},
  author = {\ifmmode \mbox{\c{S}}\else \c{S}\fi{}a\ifmmode \mbox{\c{s}}\else \c{s}\fi{}\ifmmode \imath \else \i \fi{}o\ifmmode \breve{g}\else \u{g}\fi{}lu, Ersoy and Galanakis, Iosif and Friedrich, Christoph and Bl\"ugel, Stefan},
  journal = {Phys. Rev. B},
  volume = {88},
  issue = {13},
  pages = {134402},
  numpages = {10},
  year = {2013},
  month = {Oct},
  publisher = {American Physical Society},
  doi = {10.1103/PhysRevB.88.134402},
  url = {https://link.aps.org/doi/10.1103/PhysRevB.88.134402}
}

@article{cullity2011introduction,
  author = {B. D. Cullity and C. D. Graham},
  title = {Introduction to Magnetic Materials},
  edition = {2nd},
  publisher = {Wiley},
  address = {Hoboken, NJ},
  year = {2011},
  doi = {10.1002/9780470386323}
}

@article{jena2022strain,
  author = {A. K. Jena and S. K. Mallik and M. C. Sahu and S. Sahoo and A. K. Sahoo and N. K. Sharma and J. Mohanty and S. K. Gupta and R. Ahuja and S. Sahoo},
  title = {Strain-mediated ferromagnetism and low-field magnetic reversal in Co-doped monolayer WS$_2$},
  journal = {Sci. Rep.},
  volume = {12},
  pages = {2593},
  year = {2022},
  doi = {10.1038/s41598-022-06346-w}
}

@article{zhang2021strain,
author = {Zhang, Zhi and Liu, Er and Lu, Xianyang and Zhang, Wen and You, Yurong and Xu, Guizhou and Xu, Zhan and Wong, Ping Kwan Johnny and Wang, Yichuan and Liu, Bo and Yu, Xiaojiang and Wu, Jing and Xu, Yongbing and Wee, Andrew Thye Shen and Xu, Feng},
title = {Strain-Controlled Spin Wave Excitation and Gilbert Damping in Flexible Co2FeSi Films Activated by Femtosecond Laser Pulse},
journal = {Adv. Electron. Mater.},
volume = {31},
number = {13},
pages = {2007211},
doi = {https://doi.org/10.1002/adfm.202007211},
url = {https://advanced.onlinelibrary.wiley.com/doi/abs/10.1002/adfm.202007211},
year = {2021}
}

@article{luo2018electronic,
author = {Min Luo and Chentao Yin},
title = {Electronic and magnetic properties of Al-doped WS2 monolayer under strain},
journal = {Ferroelectrics},
volume = {531},
number = {1},
pages = {114--121},
year = {2018},
publisher = {Taylor \& Francis},
doi = {10.1080/00150193.2018.1497417},

}

@article{hirohata2020review,
  author = {A. Hirohata and K. Yamada and Y. Nakatani and I.-L. Prejbeanu and B. Di{\'e}ny and P. Pirro and B. Hillebrands},
  title = {Review on spintronics: Principles and device applications},
  journal = {J. Magn. Magn. Mater.},
  volume = {509},
  pages = {166711},
  year = {2020},
  doi = {10.1063/1.4893157}
}

@article{gao2018strain,
doi = {10.1209/0295-5075/123/17002},
url = {https://doi.org/10.1209/0295-5075/123/17002},
year = {2018},
month = {aug},
publisher = {EDP Sciences, IOP Publishing and Società Italiana di Fisica},
volume = {123},
number = {1},
pages = {17002},
author = {Gao, G. Q. and Jin, C. and Zheng, W. C. and Pang, X. and Zheng, D. X. and Bai, H. L.},
title = {Strain-mediated magnetic properties of epitaxial cobalt ferrite thin films on flexible muscovite},
journal = {Europhys. Lett.},

}

@article{gueye2014bending,
    author = {Gueye, M. and Wague, B. M. and Zighem, F. and Belmeguenai, M. and Gabor, M. S. and Petrisor, T., Jr. and Tiusan, C. and Mercone, S. and Faurie, D.},
    title = {Bending strain-tunable magnetic anisotropy in Co$_2$FeAl Heusler thin film on Kapton},
    journal = {Appl. Phys. Lett.},
    volume = {105},
    number = {6},
    pages = {062409},
    year = {2014},
    month = {08},
    issn = {0003-6951},
    doi = {10.1063/1.4893157},
    url = {https://doi.org/10.1063/1.4893157},
    
}

@article{mahfouzi2020magnetoelastic,
  title = {Magnetoelastic and magnetostrictive properties of ${\mathrm{Co}}_{2}X\mathrm{Al}$ Heusler compounds},
  author = {Mahfouzi, Farzad and Carman, Gregory P. and Kioussis, Nicholas},
  journal = {Phys. Rev. B},
  volume = {102},
  issue = {9},
  pages = {094401},
  numpages = {9},
  year = {2020},
  month = {Sep},
  publisher = {American Physical Society},
  doi = {10.1103/PhysRevB.102.094401},
  url = {https://link.aps.org/doi/10.1103/PhysRevB.102.094401}
}

@article{martinez2026tailoring,
author = {Martinez Outomuro, Pablo and Navas, David and Belloso, Cantia and Lopez-Polin, Guillermo and Asenjo, Agustina},
title = {Tailoring the Anomalous Nernst Effect of Co/Pt Multilayers Grown on Strained Flexible Substrates},
journal = {Adv. Electron. Mater.},
volume = {12},
number = {11},
pages = {e70423},
doi = {https://doi.org/10.1002/aelm.70423},
url = {https://advanced.onlinelibrary.wiley.com/doi/abs/10.1002/aelm.70423},
year = {2026}
}

@article{Oh2026,
  author  = {Oh, Minsun and Kim, Yubin and Ha, Minjeong},
  title   = {Strain-decoupled magnetism in flexible spintronic sensors},
  journal = {npj Spintronics},
  volume  = {4},
  number  = {1},
  pages   = {3},
  year    = {2026},
  doi     = {10.1038/s44306-025-00122-y}
}
\end{document}


\section{S1.Details of growth conditions}
CFA films of thickness 25 nm were deposited onto flexible PI substrates using DC magnetron sputtering in a high-vacuum multi-deposition chamber (manufactured by Mantis Deposition Ltd., UK) with a base pressure better than 4 × 10$^{-8}$ milibar from an Al-rich Co-Fe-Al target. Before transfer to the sputtering chamber, the PI substrates were ultrasonically cleaned for 20 minutes in Isopropanol and then blow-dried with nitrogen gas. The sample structure is the following: PI (25 $\mu$m)/Ta (15nm)/CFA (25 nm)/Cu(5 nm) as shown in Fig.1(a) in the main manuscript. The rates of deposition for CFA, Ta and Cu layers were 0.10 \AA/s, 0.11 \AA/s and 0.13 \AA/s,  respectively, where the rate was monitored by a quartz crystal(QCM) microbalance monitor. The deposited CFA films are  amorphous under the present deposition conditions.The substrate was rotated at 20 revolutions per minute (rpm) during deposition to get homogeneity in the deposited sample. A 15 nm thick Ta buffer layer was used to reduce the roughness of PI substrate, whereas the 5 nm thick Cu capping layer was used to prevent the oxidation of the top CFA layer.

\section{S2. Strain and Stress Estimation:}
\begin{figure*}[!htbp]
\centering
\includegraphics[width=0.8\textwidth]{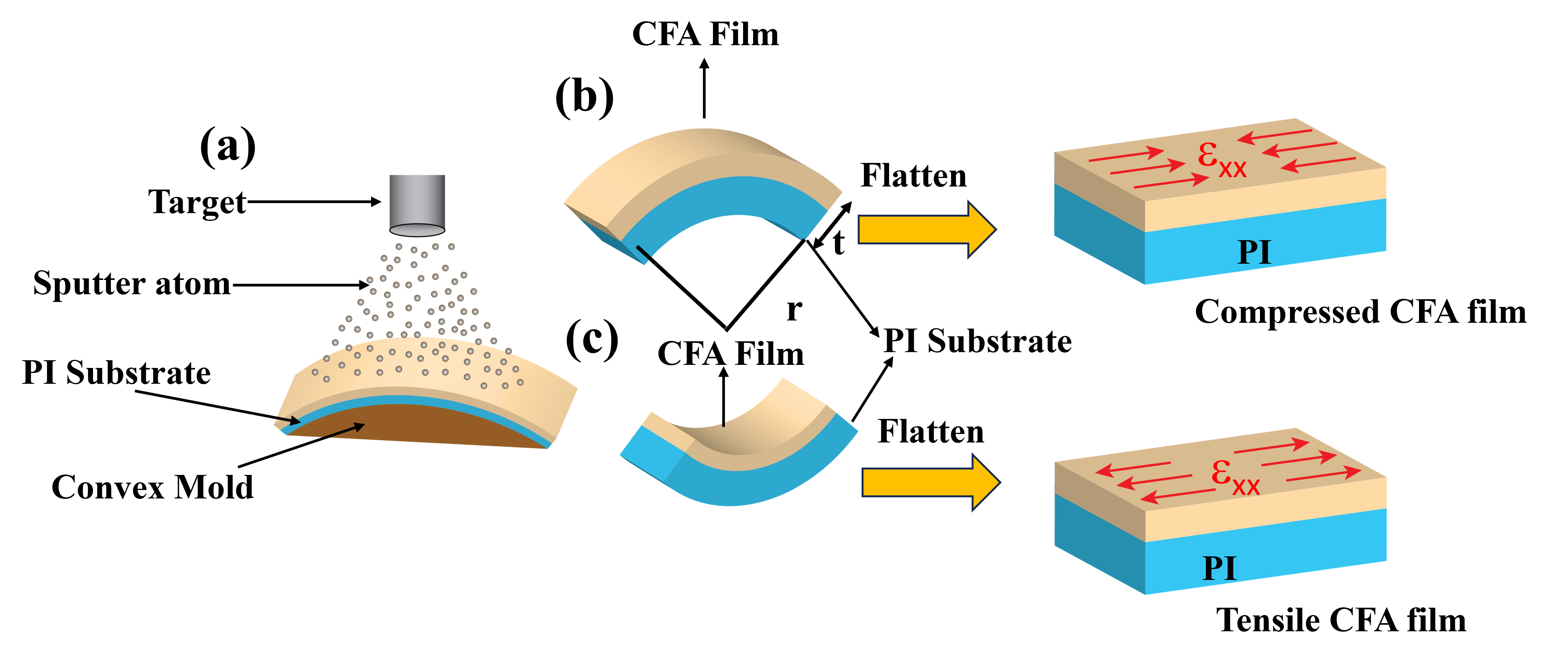}
\caption { (a) the sample deposition geometry when a PI substrate is mounted on a convex mold, (b) compressive and (c) tensile strain generation by making the as-deposited bent sample flat.}
\label{fig:Sample structure CFA film}
\end{figure*}
Mechanical strain in the flexible films was introduced through controlled bending of the polyimide (PI) substrate using molds of different radii. Strain was generated in the film using two different methods. First, the film was deposited on a flat PI substrate, and strain was later generated by sticking the sample onto convex and concave molds. 
Different tensile and compressive strains have been generated on the film by fixing it on convex and concave shaped molds, respectively. Further, magnetic properties of the sample have been recorded at both the flat as well as bend states. In the second approach for strain generation, required sample stack is deposited on flexible PI substrate which is fixed on concave and convex molds with variable radii, as shown in  Fig.~\ref{fig:Sample structure CFA film}(d). After deposition, the sample is released from the mold and flattened. Compressive strain is generated when sample is deposited on a convex-shaped flexible substrate and flattened and vice versa. Under bending, the magnetic thin film experiences either tensile or compressive  inplane strain depending on the curvature direction. For a thin film on a thick flexible substrate, the in-plane strain ($\varepsilon_{xx}$) at the film surface can be approximated as:

\begin{align}
    \mathcal{E}_T &= \frac{t}{2R+T} \\
    \mathcal{E}_C &= -\frac{t}{2R+T} 
\end{align}
$\mathcal{E}_T$ and $\mathcal{E}_C$ are the tensile and compressive strains, \textit{R} is the radius of the substrate, \textit{t} is the thickness of the sample, including the substrate thickness.
The corresponding in-plane stress ($\sigma$) is estimated using Hooke’s law under the plane-stress condition:
\begin{equation}
\sigma = \frac{E_f \, \varepsilon}{1 - \nu^2}
\end{equation}
where $E_f$ is the Young’s modulus of the magnetic film and $\nu$ is the Poisson ratio \cite{tang2014magneto}. For the present analysis, we use $E_f = 234$~GPa and $\nu = 0.33$, consistent with reported values \cite{gueye2014bending,mahfouzi2020magnetoelastic}.
Since the thickness of the magnetic layer is much smaller than that of the PI substrate, the applied strain is assumed to be fully transferred to the film. Furthermore, all applied deformations are within the elastic limit of the substrate, ensuring reversible strain without plastic deformation or film delamination.
The strain range investigated in this work ($\sim 0.03\%$--$0.25\%$) was controlled by varying the bending radius, enabling systematic tuning of the magnetoelastic response.

\begin{table*}[ht]
\centering
\caption{Summary of strain generation methods and experiments}
\label{tab:strain_methods}
\renewcommand{\arraystretch}{1.3}

\begin{tabular}{|p{2.5cm}|p{8.5cm}|p{3cm}|}
\hline
\multicolumn{1}{|c|}{\makecell{Sample \\ preparation \\ method}} & 
\multicolumn{1}{c|}{Details of strain generation} & 
\multicolumn{1}{c|}{Type of experiment}  \\
\hline

\multicolumn{1}{|c|}{Method 1} & 
\textbf{Post-deposition bending:} Films were deposited on flat PI substrates and subsequently bent over convex and concave molds to induce tensile and compressive strain, respectively. 
& \multicolumn{1}{c|}{\makecell{AFM, SEM, MOKE \\ microscopy}} \\
\hline

\multicolumn{1}{|c|}{Method 2} & 
\textbf{Pre-strained deposition:} Films were deposited on PI substrates fixed on convex and concave molds; subsequent flattening generated compressive (convex) and tensile (concave) strain.
& \multicolumn{1}{c|}{\makecell{SQUID-VSM \\ magnetometry}} \\
\hline
\end{tabular}
\end{table*}

\section{S3. Structural details and Phonon band structure} 
\begin{figure*}[!htbp]
\centering
\includegraphics[width=0.8\textwidth]{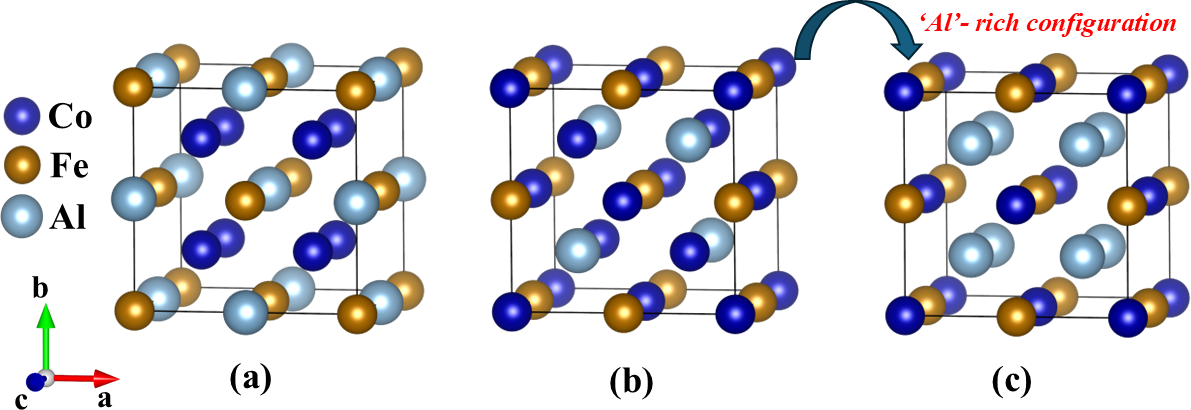}
\caption {Structural models showing the progression of chemical arrangement in Co–Fe–Al. Panel (a) represents the  ordered L2$_1$ Co$_2$FeAl lattice, while (b) represents inverse Co$_2$FeAl lattice and (c) shows the Al-rich Co-Fe-Al heusler type model.}
\label{fig:structure1}
\end{figure*}
\begin{figure*}[!htbp]
\centering
\includegraphics[width=0.5\textwidth]{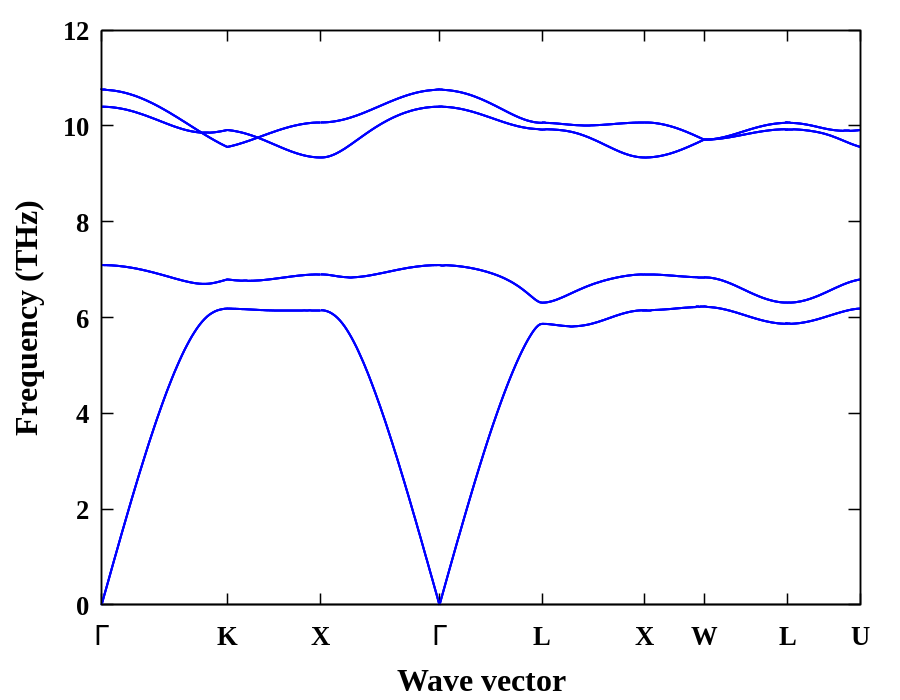}
\caption {Phonon band structure of Al$_2$CoFe along high-symmetry k-points of the Brillouin zone.}
\label{fig:structure}
\end{figure*}

To model the effect of Al enrichment, we considered both the conventional L2$_1$ full-Heusler structure of Co$_2$FeAl and an inverse-Heusler alloy (XA). Starting from a 16-atom conventional Heusler unit cell (consisting of 8-Co, 4-Fe, and 4-Si),an Al-rich environment was constructed by substituting selected Co sites in the inverse-Heusler framework with Al, yielding a nominal Al$_2$CoFe-type configuration. To emulate Al enrichment within a tractable first-principles framework, a pair of Co atoms at the ($\frac{1}{4}$, $\frac{1}{4}$, $\frac{1}{4}$) position in the inverse-Heusler alloy were replaced by Al, corresponding to a 50\% Al occupancy on the transition-metal sublattice as shown in the Fig. \ref{fig:structure1}. This configuration captures the essential local chemical environment of the experimentally realized Al-rich films (Al ~ 52 at.\%) while preserving a periodic unit cell suitable for systematic strain analysis. This approach captures the local chemical coordination expected in Al-rich sputtered films while retaining a well-defined crystalline reference structure suitable for first-principles analysis. All atomic positions and lattice parameters were fully relaxed prior to applying strain. The details of optimized interatomic  distances between Co-Fe, Co-Al and Fe-Al atomic pairs for both the structures are given in Table \ref{tab:interatomic_distances}. We have computed phonon dispersion curves within density functional peturbation theory (DFPT) as implemented in VASP using PHONOPY code \cite{phonopy-phono3py-JPCM, phonopy-phono3py-JPSJ}. The absence of imaginary frequencies confirms the dynamical stability of Al-rich phase, Al$_2$CoFe.
\begin{table}[ht]
\centering
\caption{Optimized interatomic distances (\AA) between Co–Fe, Co–Al, and Fe–Al atomic pairs for the stoichiometric Co$_2$FeAl phase and the Al-rich Al$_2$CoFe structure.}

\label{tab:Bond_lenths}
\begin{tabular}{ccccc}
\hline
Compound & Co-Fe (\AA) & Co-Al (\AA) & Fe-Al (\AA) \\
\hline
 Co$_2$FeAl &  2.46 \AA & 2.46 \AA & 2.85 \AA \\
 Al$_2$CoFe &  2.86 \AA & 2.47 \AA & 2.47 \AA \\
\hline
\label{tab:interatomic_distances}
\end{tabular}
\end{table}



\section{S4. Surface morphology, topography and compositional analysis}
\begin{figure*}[!htbp]
\centering
\includegraphics[width=0.85\textwidth]{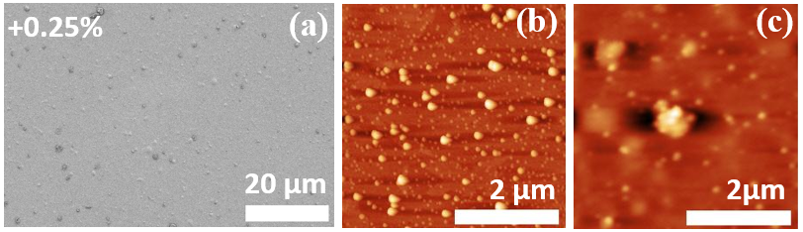}
\caption{(a) SEM image of CFA film with application of tensile 0.25\% of strain.AFM images show the topography (5×5 µm$^2$) image for (b) Si/SiO$_2$/Ta (15 nm)/CFA (25 nm)/Cu (5 nm) and (c) PI/Ta (15 nm)/CFA (25 nm)/Cu (5 nm).}
\label{fig:AFM images}
\end{figure*}
The surface morphology is studied by scanning electron microscope (SEM), as shown in Fig.~\ref{fig:AFM images}(a). The absence of cracks, delamination, or bubbles under 0.25\% tensile strain suggests that the PI/Ta(15 nm)/CFA(25 nm)/Cu(5 nm) film adheres strongly to the PI substrate, confirming good mechanical compatibility. The uniform gray contrast implies homogeneous grain distribution and no large voids, which is crucial for reliable spintronic transport because scattering centers (e.g., defects, pinholes) are minimized. Since SEM is sensitive to morphology at the micron scale, the fact that no defects appear means the strain is accommodated elastically by the thin film without plastic deformation at this strain level. Further, AFM images reveal that the surface topography remains largely unaffected when metals are deposited on a flexible PI substrate compared to a thermally oxidized Si substrate, as shown in Fig.~\ref{fig:AFM images} (b) and (c). The root mean square (RMS) roughness ($R_q$) values of the Ta(15 nm)/CFA(25 nm)/Cu(5 nm) films deposited on rigid Si/SiO$_2$ and flexible PI substrates are 3.42 nm and 3.66 nm, respectively. This indicates that the flexible PI substrate provides a sufficiently smooth platform, comparable to the conventional Si substrate, thereby ensuring uniform film growth. Such a smooth morphology is particularly important for maintaining the magnetic and electrical properties of the films.

These results confirm that the structural quality and surface morphology remain largely unaffected by both substrate type and mechanical deformation, ensuring that the observed changes in magnetic properties originate from intrinsic strain effects rather than morphological variations.

\begin{figure}[!htbp]
\centering
\includegraphics[width=0.85\textwidth]{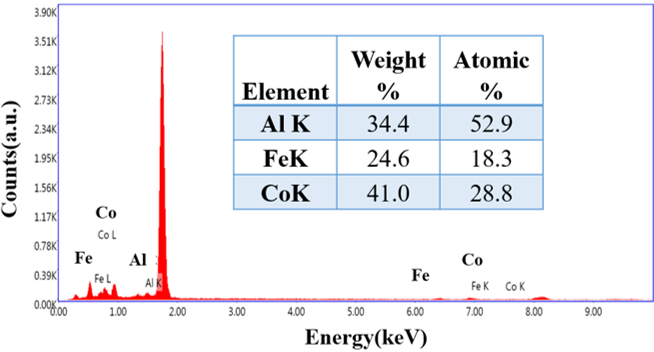}
\caption { EDXS spectrum of CFA. Table in the inset of the figure shows the atomic percentage of the constituent elements present in the sample  }
\label{fig:EDAX}
\end{figure}
Further, the elemental composition of the films was analyzed using energy-dispersive X-ray spectroscopy (EDXS) performed with the SEM, at multiple regions across the sample surface. The measurements confirm an Al content of approximately $\sim 52$~at.\%. It is also observed that the composition is uniform across different regions of the thin film surface that suggests homogeneous film growth without significant spatial variation.

\section{S5. Additional MOKE Measurements}
Additional magneto-optical Kerr effect (MOKE) measurements were performed to systematically examine the strain-dependent magnetic behavior of the films. Hysteresis loops were recorded under multiple values of both tensile and compressive strains  and along different in-plane field orientations.
As mechanical bending strain gives stress-induced anisotropy in the sample, we vary the strain from 0.03 \% to 0.25 \% by varying the radius of the mould. The magnetoelastic energy of a thin film can be expressed as EME = 3/2($\lambda \sigma$ $\sin ^2$ $\theta$), where $\lambda$ is the co-efficient of Magnetostriction, $\sigma$ is the applied stress, $\theta$ is the angle between the stress and magnetization direction.\cite{cullity2011introduction} When the product of $\lambda$ and $\sigma$ is negative, the system’s energy is minimized by creating an easy magnetic axis perpendicular to the stress axis. On the other hand, when the product is positive, the easy axis is parallel to the stress axis.   Consequently, the strength of the inverse magnetostrictive effect varies depending on the type of strain acting on the film. 
\begin{figure*}[!ht]
\centering
\includegraphics[width=0.95\textwidth]{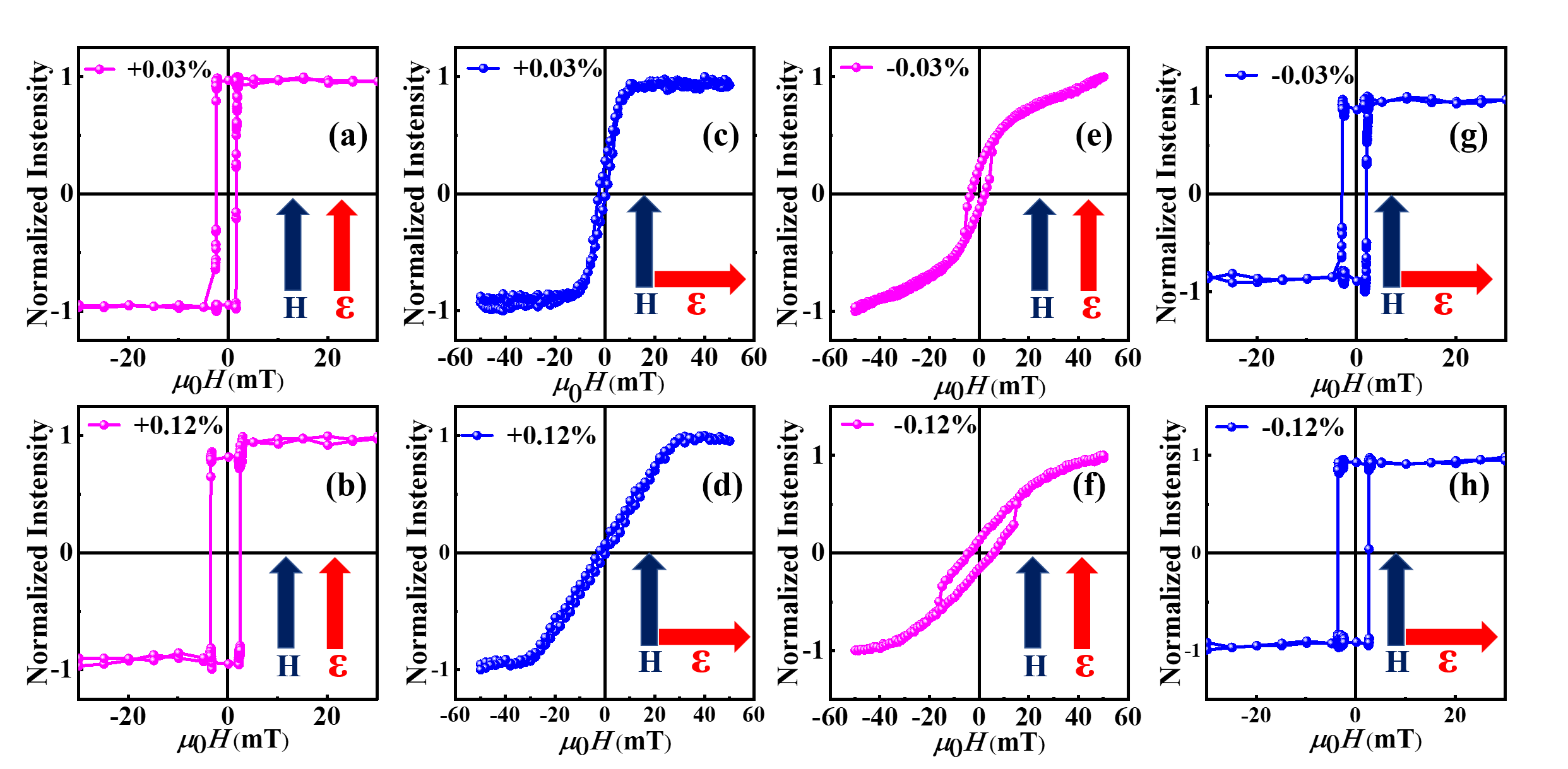}
\caption {MOKE hysteresis loops of PI/CFA films measured under different in-plane strain conditions. (a$-$d) Tensile strain (+0.03\% and +0.12\%) with the magnetic field applied parallel (a) and (b) and perpendicular (c) and (d) to the stress axis, respectively. (e$-$h) Compressive strain ($-$0.03\% and $-$0.12\%) for the magnetic field applied parallel (e) and (f) and perpendicular (g) and (h) to the stress axis. A clear evolution from square to slanted loops indicates strain-induced modification of magnetic anisotropy, including switching between easy$-$ and hard$-$axis configurations.}
\label{fig:hysteresis}
\end{figure*}
When the applied magnetic field and the stress axis are parallel, we observe a change in coercivity under tensile strain, which is shown in Fig.~\ref{fig:hysteresis} (a) and (b) for strain values of 0.03\% and 0.12\%, respectively. As the strain increases, the coercivity also increases, indicating magnetic hardening in the system. (0\% and0.25\% has shown in the main manuscript)

Further, when compressive strain is applied, the stress axis aligns with the magnetic field, causing a remarkable transformation. In this scenario, the easy axis of magnetization in the CFA film experiences a major shift, reorienting itself to become the hard axis, thus fundamentally altering the magnetic characteristics of the material as shown in Fig.~\ref{fig:hysteresis} (e) and (f). As the strain varies from -0.03\% to -0.12\%, the saturation field is also increasing. In a similar fashion, when the magnetic field is applied perpendicular to the stress axis in the case of the tensile strain varying from 0.03-0.12\%,  the magnetization requires a relatively stronger field to saturate as shown in Fig.~\ref{fig:hysteresis} (c) and (d). This magnetic hardening increases even more as the strain increases, demonstrating a clear coupling between the mechanical deformation and the film’s magnetic properties. However, under compressive strain, when the magnetic field is applied perpendicular to the stress axis, the hard axis of magnetization in the CFA film can be reoriented to become the easy axis, effectively reversing the magnetic behavior as shown in Fig.~\ref{fig:hysteresis} (g) and (h). This reveals the significant impact of strain and magnetic field orientation on CFA films. In particular, the evolution of the hysteresis shape with strain clearly demonstrates the progressive magnetic hardening under tensile strain and the corresponding reorientation of the easy axis under compressive strain.
Representative MOKE loops for selected strain values are shown in Fig.~S6, while the complete dataset is provided to illustrate the systematic evolution of the magnetic response with strain.
\section{S6. Magnetic domains under different strains in CFA films:}
\begin{figure*}[!htbp]
\centering
\includegraphics[width=0.95\textwidth]{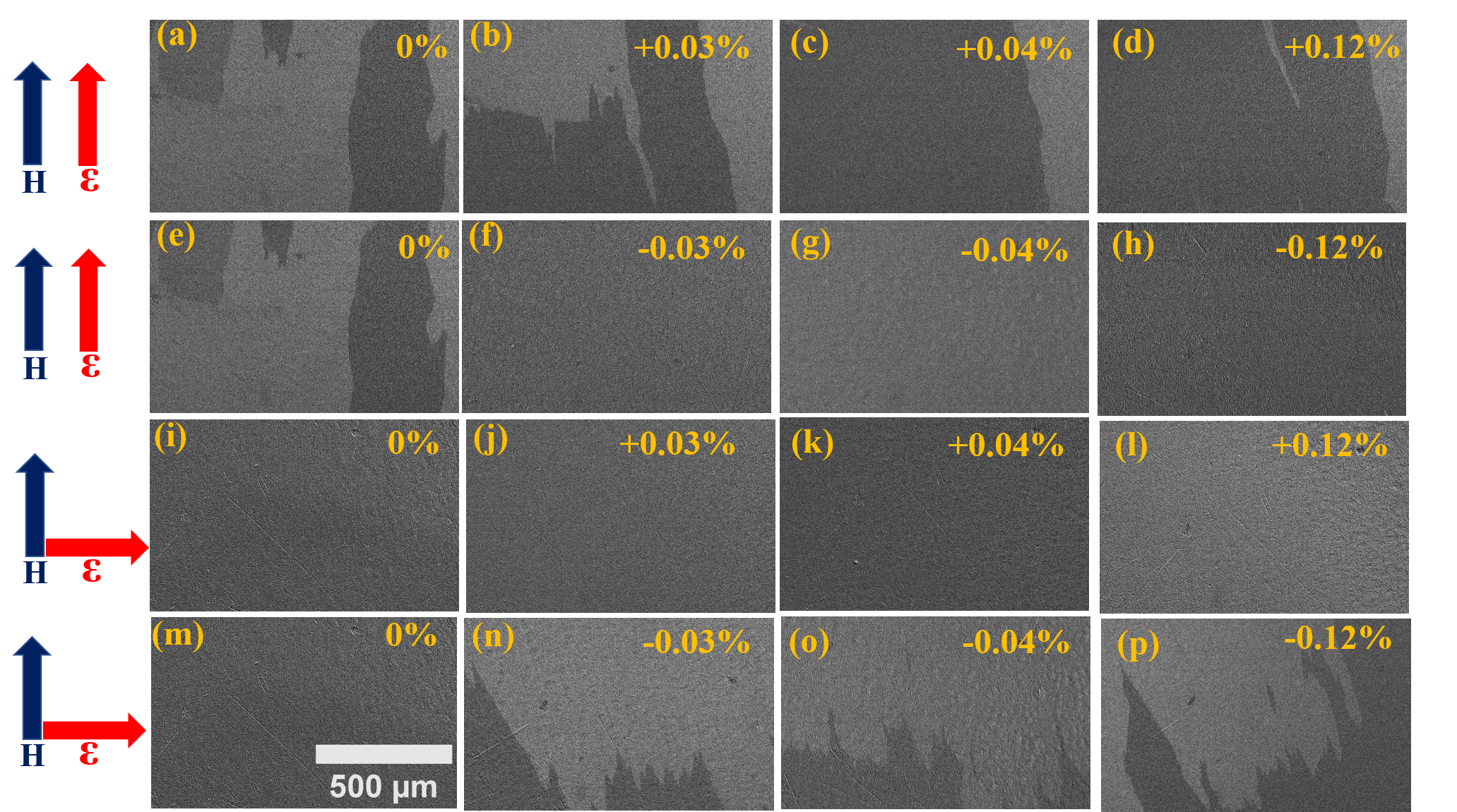}
\caption {Domain images near coercivity for tensile strain when field is applied parallel to stress axis (a-d), for compressive strain when field is applied parallel to stress axis (e-h), for tensile strain when field is applied perpendicular to stress axis (i-l), and for compressive strain when field is applied perpendicular to stress axis (m)-(p).}
\label{fig:strain_domain}
\end{figure*}
Magnetization reversal in all samples was investigated through magnetic hysteresis loop measurements, complemented by real-time domain imaging using MOKE microscope in longitudinal configuration. The analysis presented here emphasizes the role of strain-induced inverse magnetostrictive effects in governing domain nucleation and propagation. For consistency, all domain images were obtained from the same sample location, allowing direct comparison between strained and unstrained states. Representative domain patterns at coercivity for different strain conditions are shown in Fig. S5. When the external magnetic field is applied parallel to the stress axis, stripe-like domains emerge in films subjected to tensile strain, indicating that tensile strain reduces the energy barrier for domain wall nucleation and expansion (Fig.~\ref{fig:strain_domain} (a-d)). With increasing tensile strain, these nucleated dark domains grow larger, consistent with enhanced magnetoelastic anisotropy favoring domain extension along the strain axis. In contrast, films under compressive strain exhibit predominantly coherent magnetization rotation without the formation of distinct domains (Fig.~\ref{fig:strain_domain} (e-h)). When the magnetic field is applied perpendicular to the strain axis, no domain nucleation is observed in the tensile-strained films, confirming that this direction becomes magnetically hard (Fig.~\ref{fig:strain_domain} (i-l)). Conversely, in the compressively strained case, no clear domain features are detected in the unstrained (0 \%) state (Fig. 5(m)). 
However, with increasing compressive strain, domain nucleation is observed near the coercive field [(Fig.~\ref{fig:strain_domain} (n-p)], in contrast to tensile strain [(Fig.~\ref{fig:strain_domain} (b-d)], indicating a strong anisotropic effect of strain on magnetization reversal.
\section{S7. SQUID Measurements and Anisotropy Energy Estimation}
\begin{figure}[!htbp]
\centering
\includegraphics[width=0.95\textwidth]{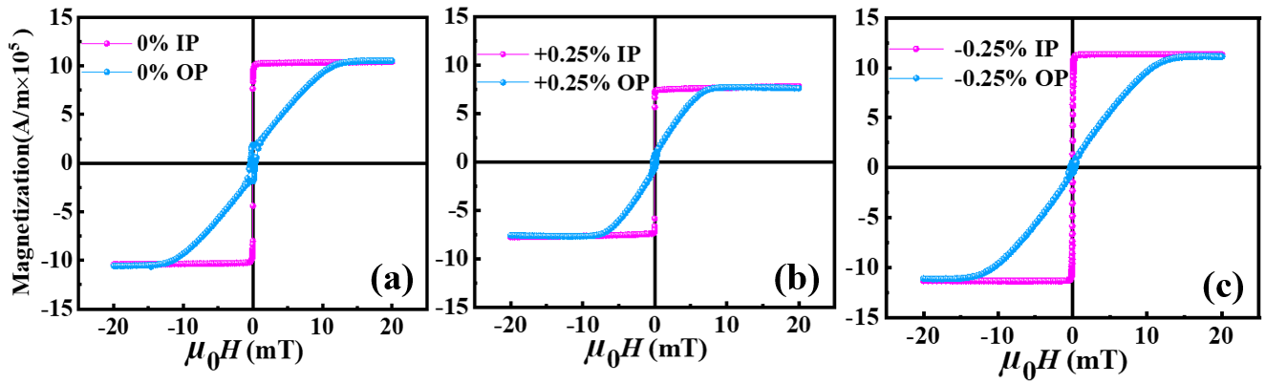}
\caption {In-Plane and Out of plane SQUID hysteresis loop is plotted, where it indicates that the sample has in-plane anisotropy for (a) unstrained sample, (b) 0.25 \% tensile strain, (c) 0.25 \% compressive strain}
\label{fig:SQUID}
\end{figure}
The in-plane and out-of-plane $M$–$H$ loops of the CFA films were measured using SQUID-VSM to probe the magnetic anisotropy and its evolution under strain. The measurements were carried out on films deposited on convex and concave molds, which were subsequently flattened to impose controlled tensile and compressive strain.
The measurements confirm that the films exhibit predominantly in-plane magnetic anisotropy, as evidenced by the lower saturation field and higher remanence in the in-plane configuration compared to the out-of-plane direction. The full set of hysteresis loops under tensile and compressive strain is provided in Fig.~S6.

The uniaxial magnetic anisotropy constant ($K_u$) was estimated from the SQUID data using the \cite{hirohata2020review}:
\begin{equation}
K_u = \frac{\mu_0 H_s M_s}{2},
\end{equation}
where $H_s$ is the saturation field along the hard axis and $\mu_0$ is the permeability of free space. The values of $H_s$ were extracted from the high-field region of the hysteresis loops.

The variation of $K_u$ with applied strain is consistent with the magnetoelastic behavior discussed in the main text. These results further support the strain-induced modulation of magnetic anisotropy and validate the extraction of magnetostriction from the experimental data.
\section{S8. Magnetostriction coefficent calculation:}

For calculating saturation magnetization, we have first subtracted the substrate contribution for different strains and unstrained states.
The uniaxial magnetic anisotropy constant ($K_u$) can be obtained from the relation given in \ref{eq:Ku}.
\begin{equation}
    K_u = \frac{1}{2} \mu_0 M_S H_S
    \label{eq:Ku}
\end{equation}
Where, $M_S$ is the saturation magnetization, and $H_S$ is saturation field along the hard axis.
The applied stress can be calculated by following the relation given in equation \ref{eq:Sigma}
\begin{equation}
    \sigma = \epsilon \frac{E_f}{1-\nu^2} 
    \label{eq:Sigma}
\end{equation}
Where $E_f$ is the Young's modulus, $\nu$ is the Poisson's ratio, and $\sigma$ is the stress. Young's modulus of CFA is 234 GPa and $\nu$ is 0.3.
From the inverse magnetostriction relation between $K_u$ and $\lambda_s$ (equation \ref{eq:Ku2}), where $\lambda_s$ is the saturation magnetostriction coefficient.
\begin{equation}
    K_u = \frac{3}{2} \lambda_s \sigma
    \label{eq:Ku2}
\end{equation}
\begin{equation}
    \lambda_S = \frac{2}{3\sigma} K_u
    \label{eq:lambda}
\end{equation}
\newpage

\section{S9: Partial Density of states}

\begin{figure*}
\centering
\includegraphics[width=0.9\textwidth]{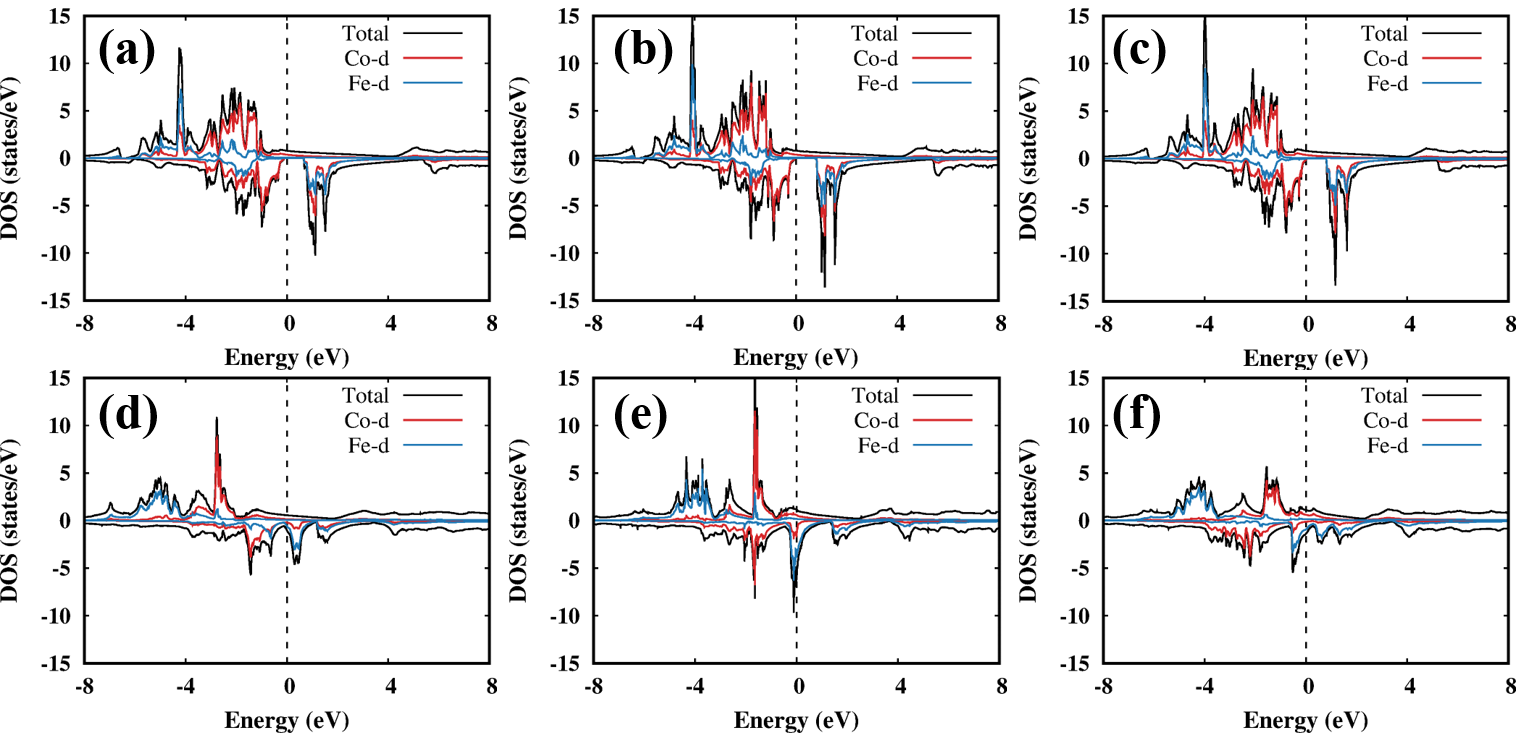}

\caption{Spin-resolved density of states (DOS) for Co$_2$FeAl (a--c) and Al$_2$CoFe (d--f) under epitaxial strain. Panels (a,d), (b,e), and (c,f) correspond to $-3\%$, $0\%$, and $+3\%$ strain, respectively. The total DOS (black) is shown together with the Co-$d$ (red) and Fe-$d$ (blue) projected contributions; positive (negative) DOS denotes the majority (minority) spin channel. The vertical dashed line marks the Fermi level $E_F=0$~eV. Compared to stoichiometric Co$_2$FeAl, the Al-rich Al$_2$CoFe exhibits pronounced strain sensitivity of the near-$E_F$ $d$-states, consistent with the strain-driven change in sublattice alignment and the resulting reduction of the net moment under tensile strain.}
\label{fig:DOS}
\end{figure*}


\section{S9: Additional details of the two-sublattice Landau interpretation}

The main manuscript now contains the essential two-sublattice Landau model and the magnetic-stiffness argument used to distinguish the weak strain response of regular \(L2_1\)-Co$_2$FeAl from the strong response of the Al-rich inverse-Heusler-like configuration. Here we retain only supplementary details concerning the interpretation and limitations of that model.

The DFT results show that the anomalous strain dependence is not produced by a uniform reduction of all local moments. The Fe moment remains positive throughout the investigated strain range, whereas the Co moment is much more sensitive and changes sign between the compressive and near-zero-strain regimes. Consequently, the decrease of the total moment under tensile strain is most naturally interpreted as an enhancement of compensation between inequivalent magnetic sublattices.

For clarity, the magnetic state may be expressed in terms of net and compensating combinations
\begin{equation}
M=m_{\mathrm{Co}}+m_{\mathrm{Fe}},\qquad
L=m_{\mathrm{Fe}}-m_{\mathrm{Co}}.
\label{eq:supp_ML}
\end{equation}
Within this representation, the regular Co$_2$FeAl phase remains predominantly in the net-moment channel: the Co and Fe moments stay parallel and vary only weakly with strain. In the Al-rich system, tensile strain increases the relative importance of the compensating channel \(L\), while compressive strain favors a more parallel alignment and therefore a larger \(M\). The anomalous response is thus better described as a redistribution between net and compensating magnetic channels than as a simple weakening of local magnetism.

The low-order Landau model should not be interpreted as a quantitative fit to the calculated sublattice moments. In particular, the Co moment in Al$_2$CoFe undergoes a relatively sharp sign reversal between the compressive regime and the vicinity of zero strain, and the tensile-side evolution is not strictly monotonic. The Fe moment also changes appreciably over the same interval. Reproducing these details would require higher-order invariants, additional strain-dependent couplings, or a microscopic parametrization directly derived from the electronic structure. The purpose of the phenomenology is instead to identify the relevant soft compensation mode and to show why its strain response can be strongly enhanced when the curvature of the magnetic free-energy surface becomes small.

This interpretation is consistent with the DOS results in Fig.~\ref{fig:DOS}. In stoichiometric Co$_2$FeAl, the near-\(E_F\) electronic structure changes only weakly with strain, consistent with a stiff ferromagnetic state. In Al$_2$CoFe, the sharper strain-induced redistribution of Co- and Fe-derived spectral weight near \(E_F\) is consistent with a softer magnetic sector and with the strong variation of the Co sublattice moment. The Landau description is therefore used only as a bridge between the DFT electronic-structure results and the observed macroscopic strain response.

\bibliography{References}